\documentclass[aps,preprint,tightenlines,superscriptaddress,showpacs,floatfix,nofootinbib]{revtex4}
\usepackage{amsmath}
\usepackage{epsfig}

\newcommand{\bra}{\langle}
\newcommand{\ket}{\rangle}
\newcommand{\ct}{|\bar 3_{12}3_{34}\rangle}
\newcommand{\cs}{|6_{12}\bar 6_{34}\rangle}
\newcommand{\cuut}{|1_{13}1_{24}\rangle}
\newcommand{\cout}{|8_{13}8_{24}\rangle}
\newcommand{\cuuc}{|1_{14}1_{23}\rangle}
\newcommand{\couc}{|8_{14}8_{23}\rangle}

\newcommand{\bs}[1]{\ensuremath{\boldsymbol{#1}}}
\begin{document}

\title{Exotic meson-meson molecules and compact four--quark states}

\author{J. Vijande}
\affiliation{Departamento de F\'{\i}sica At\'{o}mica, molecular y Nuclear, Universidad de Valencia (UV)  
and IFIC (UV-CSIC), Valencia, Spain.}
\affiliation{Departamento de F\'\i sica Fundamental, Universidad de Salamanca, 
E-37008 Salamanca, Spain}
\author{A. Valcarce}
\affiliation{Departamento de F\'\i sica Fundamental, Universidad de Salamanca, 
E-37008 Salamanca, Spain}
\author{N. Barnea}
\affiliation{The Racah Institute of Physics, The Hebrew University, 91904,
Jerusalem, Israel}
\affiliation{Institute for Nuclear Theory, University of Washington, 
Seattle, WA 98195, USA}

\date{\today}
\begin{abstract}
We present an exact calculation of $S$ and $P$ wave $QQ\bar n\bar n$ states 
using different standard nonrelativistic quark--quark potentials.
We explore in detail the charm and bottom sectors looking for bound states 
that could be measured within existing facilities. Against the proliferation
of four--quark states sometimes predicted in the literature, we found a small 
number of candidates to be stable. We analyze their properties in a trial to
distinguish between compact and molecular states. Possible decay modes 
are discussed.
\end{abstract}

\maketitle

\section{Introduction}
\label{intr}

The existence of stable $QQ\bar n\bar n$ states 
has been the topic for discussion since the early 80's~\cite{Ade82}.
These states are of particular interest since they are manifestly 
exotic, i.e., heavy flavor quantum number $\pm2$ with baryon number equal 0. 
If they would lie below the threshold for dissociation into two ordinary 
hadrons they would be narrow and should show up clearly in the experimental 
spectrum. There are already estimates of the production rates
indicating they could be produced and detected at present 
(and future) experimental facilities~\cite{Fab06}.
Four--quark states seem to be necessary to tame the bewildering
landscape of new meson states. Non--exotic four--quark states
embedded in the meson spectra have been proposed as a thoughtful explanation 
of the proliferation and peculiar properties of light scalar--isoscalar
or open--charm mesons \cite{Ams04}. There are recent indications
of the possible existence of molecular four--quark states, 
suggested from the observation of the $Z^+(4430)$~\cite{Abe07}. 
A step forward scrutinizing the structure of low-energy hadrons
would be to demonstrate the existence of stable manifestly exotic multiquark
states, the $QQ\bar n\bar n$ system being an ideal candidate. 

To illustrate the theoretical landscape,
we present in Table~\ref{tres-10} a summary 
of different approaches to the manifestly exotic
four--quark spectroscopy~\cite{Zou86,Lip86,
Car88,Man93,Sil93,Pep96,Bri98,Gel03,Jan04,Vij04b,Nav07,Ebe07,Zha07}.
Exotic multiquarks were examined in Ref.~\cite{Zou86} solving
the four--body problem by three different variational methods
with a nonrelativistic potential considering explicitly 
virtual meson--meson components in the wave function. 
In Ref.~\cite{Lip86} four--quark states were studied using a 
variational approach with trial wave functions
whose interaction energies are approximately given by known hadron masses.
Ref.~\cite{Car88} used a potential derived from the MIT bag model in the Born-Oppenheimer 
approximation. The calculations were done by means of the Green's function Monte Carlo method. 
$b\bar n-b\bar n$ molecules loosely bound by the one-pion exchange were obtained 
in Ref.~\cite{Man93} using chiral perturbation theory.
Ref.~\cite{Sil93} analyzed $L=0$ four--quark systems with   
the Bhaduri potential and a variational method in a harmonic oscillator basis up to $N=8$
quanta. Ref.~\cite{Pep96} discussed the stability of multiquark systems using different parametrizations of
a Goldstone boson exchange model and a variational formalism with gaussian trial wave functions.
In Ref.~\cite{Bri98} the Bhaduri potential was reexamined by means of a variational method that allows nonzero 
internal orbital angular momentum in the subsystems of quarks and antiquarks.
The existence of a shallow tetraquark state $cc\bar u\bar d$ was 
discussed in Ref.~\cite{Gel03} using semiempirical
mass relations. Ref.~\cite{Jan04} designed a powerful method, similar to the stochastic variational 
approach~\cite{Var95}, accommodating two free-meson asymptotic states. It was applied to the
Bhaduri potential, fixing the results of Refs.~\cite{Sil93,Bri98}. Ref.~\cite{Vij04b} 
analyzed multiquark states with a variational 
formalism using gaussian trial wave functions with only quadratic terms in the Jacobi coordinates. 
QCD sum rules were used in Ref.~\cite{Nav07} to 
study the possible existence of an axial diquark-antidiquark bound state. In Ref.~\cite{Ebe07} the ground 
state of tetraquarks were evaluated assuming a diquark-antidiquark structure, 
reducing the relativistic four--body problem to the solution of two relativistic 
two-body problems. Ref.~\cite{Zha07} discussed the possible existence of four--quark bound states 
within the framework of the chiral $SU(3)$ quark model by means of a variational approach using gaussian 
trial wave functions.

As seen in Table~\ref{tres-10}, there is a remarkable agreement on the 
existence of a $I=0$, $J^P=1^+$ $bb\bar n\bar n$ bound state
and also, although not so neatly, on the existence of a $cc\bar n\bar n$ one (along this work $n$
stands for a light $u$ or $d$ quark). However, among the different theoretical approaches
only a few had paid attention to other quantum numbers,
trying to elucidate if a proliferation of four-quark states is predicted.
Hence, in this work our purpose will be twofold. On the one hand,
we shall try to shed some light on this topic by making 
a detailed analysis of the $QQ\bar n\bar n$ spectra 
discussing each set of quantum numbers.
On the other hand, the Achilles' heel of almost all the approaches described above is the lack of an exact numerical 
method to solve the four--body problem. Most of these works rely on variational calculations with different 
types of trial wave functions or on semiempirical mass relations. The
importance of this problem requires numerical methods able
to provide exact solutions with controlled numerical uncertainties.

To this end we shall use a
new approach based on the hyperspherical formalism recently developed
to solve exactly the four--quark problem \cite{Bar06,Vij07}. 
The idea is to perform an expansion of the trial wave function in terms of the
hyperspherical harmonics (HH) functions. This allows to generalize the simplicity of the spherical harmonic
expansion for the angular functions of a single particle motion to a system of
particles by introducing a global length $\rho$, the hyperradius, and a
set of angles, $\Omega$. For the HH expansion to be practical, the
evaluation of the potential energy matrix elements must be feasible. The main
difficulty of this method is to construct HH functions with proper
permutational symmetry for a
system of identical particles. This is a difficult problem that may be
overcome by means of the HH formalism based on the symmetrization of the
$N-$body wave function with respect to the symmetric group using the Barnea and
Novoselsky algorithm~\cite{Nir9798}. 
For systems containing only two pairs of identical particles this
problem is greatly simplified making the appropriate choice of Jacobi coordinates.
In Ref.~\cite{Vij07} we have developed the HH formalism for the $Q\bar Q n\bar
n$ system, and we will use it here with the appropriate modifications to study the
exotic $QQ\bar n\bar n$ four--quark spectra.

The manuscript is organized as follows. In Sec.~\ref{tech} we briefly revise
the HH formalism for the $QQ\bar n\bar n$ system and present the quark models 
used. In Sec.~\ref{pepe} we introduce observables that may allow to 
distinguish between 
unbound and compact or molecular four-quark bound states. In 
Sec.~\ref{results} the results and the analysis of the $cc\bar n\bar n$ and $bb\bar n\bar n$
spectroscopy are presented. In Sec.~\ref{Experiment} the possibility of measuring the 
predicted bound states within current
experimental facilities is discussed. Finally, we summarize in Sec.~\ref{summary} our conclusions.

\section{Technical details}
\label{tech}

Within the HH expansion, the four--quark wave function 
can be written as a sum of outer products 
of color, isospin, spin and configuration terms
\begin{equation}
    |\phi_{CISR}\ket= |{\rm Color}\ket |{\rm Isospin}\ket
               \left[|{\rm Spin}\ket \otimes| R \ket \right]^{J M} \, ,
\end{equation}
such that the four-quark state is a color singlet with well defined
parity, isospin and total angular momentum.
In the following we shall assume that particles $1$ and $2$ are the $Q$-quarks
and particles $3$ and $4$ are the $n$-quarks. Thus,
particles 1 and 2 are identical, and so are 3 and 4.
Consequently, the Pauli principle leads to the following 
conditions,
\begin{equation}\label{pauli}
\hat P_{12}|\phi_{CISR}\ket=\hat P_{34}|\phi_{CISR}\ket=-|\phi_{CISR}\ket \, ,
\end{equation}
$\hat P_{ij}$ being the permutation operator of particles $i$ and $j$.

Coupling the color states of two quarks (antiquarks) can yield two possible
representations, the symmetric $6$-dimensional, $6$ ($\bar 6$),
and the antisymmetric $3$-dimensional, $\bar 3$ ($3$).
Coupling the color states of the quark pair with that of the antiquark pair
must yield a color singlet. Thus, there are only two possible color states for a
$QQ\bar q \bar q$ system \cite{Jaf77},
\begin{equation}
  |{\rm Color}\ket = \{ | \bar 3_{12} 3_{34} \ket , | 6_{12} \bar 6_{34} \ket\}\,.
\end{equation}
These states have well defined symmetry under permutations, Eq. (\ref{pauli}).
The spin states with such symmetry can be obtained in the following way,
\begin{equation}
  |{\rm Spin}\ket = |((s_1,s_2)S_{12},(s_3,s_4)S_{34})S\ket
                  = | (S_{12} S_{34}) S \ket\;.
\end{equation}
The same holds for the isospin, $|{\rm Isospin}\ket=|(i_3,i_4)I_{34} \ket$,
which applies only to the $n$-quarks, thus $I=I_{34}$.

We use the HH expansion to describe the spatial
part of the wave function. We choose for convenience the $H$-type 
Jacobi coordinates, 
\begin{eqnarray}
\bs{\eta}_1 & =&  \mu_{1,2}(\bs r_2 - \bs r_1) \, , \cr
\bs{\eta}_2 & =&  \mu_{12,34} 
                 \left(\frac{m_3\bs r_3+m_4\bs r_4}{m_{34}} 
                      -\frac{m_1\bs r_1+m_2\bs r_2}{m_{12}} \right) \, , \cr
\bs{\eta}_3 & =&  \mu_{3,4}(\bs r_4 - \bs r_3) \, ,
\end{eqnarray}
were $m_{ij}=m_i+m_j$, $\mu_{i,j}=\sqrt{m_i m_j/m_{ij}}$, and $m_{1234}=m_1+m_2+m_3+m_4$.
Using these vectors, it is easy
to obtain basis functions that have well defined symmetry under permutations
of the pairs $(12)$ and $(34)$. In the HH formalism the three Jacobi vectors 
are transformed into a single length variable,
$\rho=\sqrt{\eta_1^2+\eta_2^2+\eta_3^2}$, and $8$-angular variables, $\Omega$, that
represent the location on the $8$-dimensional sphere. 
The spatial basis states are given by 
\begin{equation}
\bra \rho \Omega |R\ket = U_{n}(\rho){\cal Y}_{ [K]} (\Omega) \, , 
\end{equation}
were ${\cal Y}_{ [K]} $ are the HH functions, and 
$[K]\equiv \{K K_{12} L M_L L_{12} \ell_3 \ell_2 \ell_1\}$.
The quantum number $K$ is the grand angular momentum, 
$L M_L$ are the usual orbital angular momentum
quantum numbers, and $\ell_i$ is the angular momentum associated
with the Jacobi vector $\bs{\eta}_i$. The quantum
numbers $K_{12}, L_{12}$ correspond to the intermediate coupling of
$\bs{\eta}_1$ and $\bs{\eta}_2$.
The Laguerre functions are used as the hyper--radial basis
functions $U_n(\rho)$.

The Pauli principle, Eq. (\ref{pauli}), leads to the following restrictions
on the allowed combinations of basis states:
\begin{itemize}
\item[{(i)}] $(-1)^{S_{12}+\ell_1}=+1$, $(-1)^{S_{34}+I+\ell_3}=-1$ 
for the $| 6_{12} \bar 6_{34} \ket$ color state,
\item[{(ii)}] $(-1)^{S_{12}+\ell_1}=-1$, $(-1)^{S_{34}+I+\ell_3}=+1$ for 
the $| \bar 3_{12} 3_{34} \ket$ state.
\end{itemize}

Assuming non-relativistic quantum 
mechanics we solve the four-body Schr\"odinger 
equation using the basis states described above. The grand angular momentum
$K$ is the main quantum number in our expansion and the
calculation is truncated at some $K$ value.

For our study we will use two standard quark potential models.
The constituent quark cluster (CQC) model was proposed in the 
early 90's in an attempt to
obtain a simultaneous description of the nucleon-nucleon
interaction and the baryon spectra \cite{Rep05}.
Later on it was generalized to all flavor sectors giving 
a reasonable description
of the meson \cite{Vij05a} and baryon spectra \cite{Vij04}. 
The model contains Goldstone boson exchange between quarks, a 
one-gluon-exchange (OGE) potential, and a screened confined
interaction as dictated by unquenched lattice calculations~\cite{Bal01}.
The model parameters have been taken from Ref.~\cite{Vij05a} with 
the exception of the OGE regularization parameter, see Ref.~\cite{Vij07} for details.
In the following we shall denote this parametrization
as CQC and the standard parametrization of Ref. \cite{Vij05a} will be referred to as 
CQC$_{18}$. Explicit expressions 
of the interacting potentials and a detailed discussion of the model can be found in 
Ref.~\cite{Vij05a}.
The Bhaduri-Cohler-Nogami (BCN) model was proposed in the early 80's in an
attempt to obtain a unified description of meson and baryon 
spectroscopy~\cite{Bha81}. It was later on applied to  
study the baryon spectra~\cite{Sil85} and four-quark
systems~\cite{Sil93}. The model retains the 
most important terms of the one-gluon exchange (OGE) interaction,
namely coulomb and spin-spin, and a linear confining potential.
The parameters are taken from Ref.~\cite{Sil93}.

A summary of the energies obtained with both models, CQC and BCN, for selected meson states 
that may appear in the thresholds of the studied four--quark systems
are given in Table~\ref{tmeson}, together with the corresponding experimental energies.

\begin{figure}[t] 
\caption{$H$--type Jacobi vectors.}
\epsfig{file=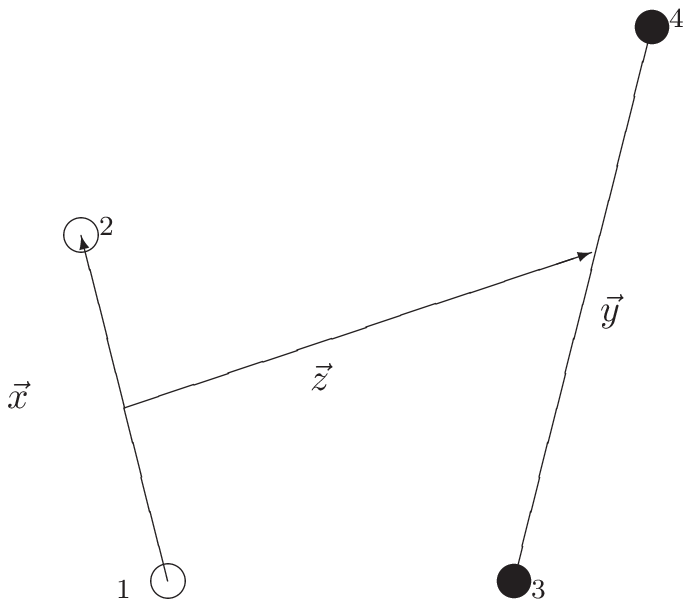, width=8cm}
\label{figH}
\end{figure}
 
\section{Bound states, compact and molecular}%
\label{sec:compact}
\label{pepe}
\subsection{Threshold determination}
\label{thre}

As thoroughly discussed in Ref.~\cite{Vij07}, in order to discriminate between four--quark bound states and
simple pieces of the meson-meson continuum, one has to carefully determine 
the two-meson states that constitute the thresholds for each set of quantum numbers.
Dealing with strongly interacting particles, the two-meson states should have well defined total angular 
momentum ($J$), parity ($P$), and a properly symmetrized wave function if two identical mesons
are considered (spin-statistics theorem). When noncentral forces are not taken into account, orbital angular 
momentum ($L$) and total spin ($S$) are also good quantum numbers. We give in Tables~\ref{th1},~\ref{th2},
~\ref{th3},~\ref{th2b}, and~\ref{th3b} 
the lowest threshold in both cases, that we will refer to 
as coupled (CO) and uncoupled (UN) schemes respectively, together with the final state relative orbital angular 
momentum of the decay products. We would like to emphasize that although we use
central forces in our calculation the CO scheme is the relevant one
for observations, since a small non-central component in the
potential is enough to produce a sizeable effect on the width of a state.
In Table~\ref{th1} we summarize the thresholds obtained using the experimental energies given in 
Ref.~\cite{PDG06}. In Tables~\ref{th2} and~\ref{th3} we quote the thresholds obtained with the CQC model for the charm 
and bottom sectors respectively, and in Tables~\ref{th2b} and~\ref{th3b} those for the BCN model.

An important property of the $QQ\bar n\bar n$ system, that is
crucial for the possible existence of bound states, 
is the fact that only one physical threshold
$(Q \bar n)(Q\bar n)$ is allowed. Consequently, particular modifications of 
the four--quark interaction, for instance a strong color-dependent 
attraction in the $QQ$ pair, would not be translated into any asymptotically free two-meson state.
As discussed in Ref.~\cite{Vij08b}, this is not a general property in the four--quark spectroscopy, 
as the $Q\bar Q n\bar n$ four--quark state has two allowed physical thresholds:
$(Q\bar Q)(n\bar n)$ and $(Q\bar n)(n\bar Q)$.

\subsection{Figures of merit}

The relevant quantity for analyzing the stability of any four--quark state is $\Delta_E$, 
the energy difference between the mass of the four--quark system and that of the 
lowest two-meson threshold,
\begin{equation}
\label{delta}
\Delta_E=E_{4q}-E(M_1,M_2)\, ,
\end{equation}
where $E_{4q}$ stands for the four--quark energy and $E(M_1,M_2)$ for the energy of the 
two-meson threshold. Thus, $\Delta_E<0$ indicates that all fall-apart decays 
are forbidden, and therefore one has a proper bound state. $\Delta_E\ge 0$ 
will indicate that the four--quark solution corresponds to 
an unbound threshold (two free mesons).

\begin{center}
\begin{figure}[t]
\caption{$E_{4q}/E(M_1,M_2)$ as a function of $K$ for the $J^P(L,S,I)=1^+(0,1,0)$ (solid line),
$0^+(0,0,0)$ (dashed line), and $1^-(1,0,1)$ (dash-dotted line) with the CQC model.}
\label{f1}
\vspace*{-1.0cm}
\hspace*{-0.7cm}
\epsfig{file=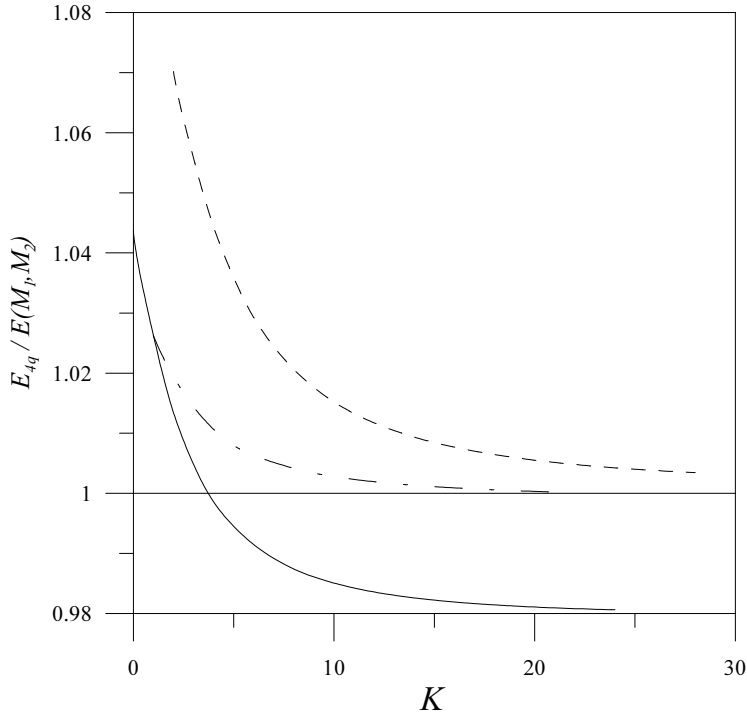, width=10cm}
\end{figure}
\end{center}

One of the main difficulties in studying four--quark states, already discussed 
in Refs.~\cite{Bar06,Vij07}, is the slow convergence of unbound solutions. 
We show in Fig.~\ref{f1} the evolution of 
$E_{4q}/E(M_1,M_2)$ as a function of $K$ for three
different states, two of them bound (solid and dash-dotted lines)
and one unbound (dashed line). Although all of them converge 
for large enough values of $K$, the correct description of the two-meson threshold for
unbound states is slow and time consuming.   
A helpful tool to minimize this problem was proposed in Ref.~\cite{Bar06}
through an extrapolation of the four--quark energy using the expression
\begin{equation}
\label{extra}
E_{4q}(K)=E_{4q}(K=\infty)+{\frac{a}{K^b}}\,,
\end{equation}
where $E_{4q}(K=\infty)$, $a$ and $b$ are fitted parameters.
When this extrapolation is used
for unbound states, one can observe how the four--quark energies 
reproduce the thresholds to within a few MeV. 

A second important quantity to characterize the possible existence of a bound state
is the evolution of the root mean square radius as a function of $K$. While for
unbound states, the components of the four-quark state will tend to be far away
when increasing $K$, for a bound state the radius should stabilize when increasing $K$.
In order to compare four--quark against two free--meson states we define
the root mean square radius for four (two) quark systems
\begin{equation}
{\rm RMS}_{4q(2q)}= \left(\frac{\sum_{i=1}^{4(2)} m_i \langle (r_i-R)^2\rangle}{\sum_{i=1}^{4(2)} m_i}\right)^{1/2}\,,
\end{equation}
and its corresponding ratio
\begin{equation}
\label{delta-r}
\Delta_R=\frac{{\rm RMS}_{4q}}{{\rm RMS}_{M_1}+{\rm RMS}_{M_2}}\,,
\end{equation}
where ${\rm RMS}_{M_1}+{\rm RMS}_{M_2}$ stands for the sum of the radii of the mesons corresponding to the lowest threshold. 
\subsection{Compact vs molecular states}

Besides trying to unravel the possible existence of
bound $QQ\bar n\bar n$ states
one should try to understand whether it is possible to differentiate between compact
and molecular states. A molecular state may be understood as a four--quark state
containing a single physical two-meson component, i.e., a unique singlet-singlet
component in the color wave function with well-defined spin and isospin quantum numbers.
One could expect these states not being deeply bound and therefore having a size
of the order of the two-meson system, i.e., 
$\Delta_R\sim1$. Opposite to that,
a compact state may be characterized by its involved structure on the
color space, its wave function containing different singlet-singlet
components with non negligible probabilities. One would expect
such states would be smaller than typical two-meson systems, i.e.,
$\Delta_R < 1$. Let us notice that while $\Delta_R>1$ but finite would 
correspond to a meson-meson molecule 
$\Delta_R\stackrel{K\to\infty}{\longrightarrow}\infty$ 
would represent an unbound threshold.

One may illustrate the situation described above by the deuteron and the
$H$-dibaryon examples. Let us try to draw the analogy between these
states and the $QQ \bar n \bar n$ system. The deuteron has a small binding
energy of
$-$2.225 MeV and a ratio between its root mean square charge radius 
(2.139 fm) and the one of two protons (1.75 fm) 
of 1.222~\cite{Cod05}, $\Delta_R=1.22$ and $\Delta_E=-$2.225 MeV in our notation. 
Should the deuteron be considered as a pure baryon-baryon molecule? 
Although Ref.~\cite{Jaf07} emphasized the difficulties to identify
pure hadron-hadron molecules close to thresholds, in the deuteron case
it was long-ago justified~\cite{Wei65}. The 
probability of physical two-baryon states other than nucleon-nucleon
is meaningless~\cite{Rep05}. Therefore it constitutes a clear example
of a molecular state. The postulated $H$-dibaryon would however fit
into the picture of compact states, its wave function presenting relevant
components of different singlet-singlet physical channels: $\Lambda\Lambda$,
$N\Xi$, and $\Sigma\Sigma$ at least~\cite{Sak99}. 

This last aspect makes contact with the role played by hidden-color configurations,
color singlets built by nonsinglet constituents. 
There are three different ways for coupling two quarks and two antiquarks 
in a colorless state, 
\begin{eqnarray}
[(q_1q_2)(\bar q_3\bar q_4)]&=&\{\ct,\cs\}\cr
[(q_1\bar q_3)(q_2\bar q_4)]&=&\{\cuut,\cout\}\cr
[(q_1\bar q_4)(q_2\bar q_3)]&=&\{\cuuc,\couc\} \, .
\end{eqnarray}
Each coupling scheme allows to 
define a color basis where the four--body problem can be solved. The first basis, 
$[(q_1q_2)(\bar q_3\bar q_4)]$, being the most suitable one to deal with the Pauli principle, 
is made entirely by hidden--color vectors. The other two are hybrid basis that contain 
both singlet--singlet (physical) and octect--octect (hidden--color) components. 
It is possible to prove from simple group theory arguments that once we have solved the four--body 
problem for a system composed of two identical quarks ($QQ$) and two identical antiquarks 
($\bar n\bar n$), there is a minimum value 
for the octect--octect component probability of the wave function either in the 
$[(Q_1\bar n_3)(Q_2\bar n_4)]$ or the $[(Q_1\bar n_4)(Q_2\bar n_3)]$ couplings: 
$P_{88}^{13,24},P_{88}^{14,23}\in[1/3,2/3]$. 
It can also be proved that for a four-quark threshold state 
$P_{88}^{13,24}=P_{88}^{14,23}=4/9$. 
Does this imply an important hidden--color 
component in all $QQ\bar n\bar n$ states? The answer is no. In Ref.~\cite{Ped87} it was proved 
than any physical state can be expanded in terms of a basis constructed by direct product 
of mesonic and/or baryonic states not necessarily linearly independent. 
Therefore one can express any $QQ\bar n\bar n$ state in terms of the singlet--singlet 
component of the $[(Q_1\bar n_3)(Q_2\bar n_4)]$ and $[(Q_1\bar n_4)(Q_2\bar n_3)]$ basis. 

This discussion can be made more quantitative. Let us assume that $\{P,Q\}$ and $\{\hat P,\hat Q\}$
are the projectors associated to two orthonormal basis that are not orthogonal to each
other, i.e., $P\hat P \mid\phi\rangle \ne 0$ and $P\hat Q \mid\phi\rangle \ne 0$ 
for an arbitrary state $\mid\phi\rangle$.
This would be the case of the two orthonormal basis: $\{\cuut,\cout\}$ and $\{\cuuc,\couc\}$.
Any arbitrary state can be written as
\begin{equation}
\mid\Psi\rangle \, = \, P\mid\Psi\rangle \, + \, Q\mid\Psi\rangle \, ,
\label{ii}
\end{equation}
and the probability of the state associated to $P$ or $\hat P$ will be given by~\cite{Vij08},
\begin{eqnarray}
{\cal P}^{\mid\Psi\rangle}({[u]})&=&\frac{1}{2(1-\cos^2\alpha)}
\left[ \left\langle\Psi\mid P\hat Q \mid\Psi\right\rangle
+\left\langle\Psi\mid \hat Q P \mid\Psi\right\rangle\right] \cr
{\cal P}^{\mid\Psi\rangle}({[u']})&=&\frac{1}{2(1-\cos^2\alpha)}
\left[ \left\langle\Psi\mid \hat P Q \mid\Psi\right\rangle +
\left\langle\Psi\mid Q \hat P \mid\Psi\right\rangle\right] \, ,     
\label{proeq}
\end{eqnarray}
where
$P=\left.\mid u \right\rangle \left\langle u \mid\right.$ and
$\hat P=\left.\mid u' \right\rangle \left\langle u' \mid\right.$ and
${\rm cos}\alpha=\left\langle u' \mid u \right\rangle$. For a molecular
state either ${\cal P}^{\mid\Psi\rangle}({[u]})$
or ${\cal P}^{\mid\Psi\rangle}({[u']})$ would be equal to zero while
for a compact state both will be different from zero.

\section{Results}
\label{results}

\subsection{Comparison to other numerical methods}
\label{compa}

To illustrate the performance of the numerical procedure described in Sec.~\ref{tech} it is convenient to
compare with other numerical methods to understand its capability and
advantages, if any. As outlined in Sec.~\ref{intr}, in the past decades there have been several 
attempts to study multiquarks containing explicit charm 
or bottom flavors. Among them we shall analyze the calculation of the
$(L,S,I)=(0,1,0)$ $cc\bar n\bar n$ state of Refs.~\cite{Jan04,Sil93}
using the BCN potential and the results obtained in Ref.~\cite{Vij04b}
with the CQC model.

We present in Table~\ref{tres-4} results for different $L=0$ spin-isospin
$cc \bar n\bar n$ states within the CQC model. We quote in the first column the
results obtained with a variational calculation using gaussian trial wave functions 
with only quadratic terms in the Jacobi coordinates~\cite{Vij04b} ($S$ wave approximation). Such approximation is
recovered in our formalism requiring $\ell_i=0$ for all pairs. These results
are quoted in the second column up to $K=10$, being fully converged,
and reproducing exactly the variational results\footnote[1]{We have redone 
the calculation of Ref.~\cite{Vij04b} with the parameters used in this work
for a proper comparison.}.
The relevance of large relative orbital angular momenta can be judged
by looking at the last column, where almost exact HH results up to $K=24$ are given,
the difference in some cases being larger
than 200 MeV. This effect was not appreciated in Ref.~\cite{Vij04b} since 
their importance was estimated using only one gaussian for the radial wave function.

Using the variational method of Ref.~\cite{Vij05b}, where nonzero relative orbital angular momenta 
were considered, we have obtained 3861.38 MeV for the energy and 0.363 fm for the root mean square radius,
respectively, of the $(L,S,I)=(0,1,0)$ $cc\bar n\bar n$ four--quark state. This result is in perfect
agreement with that obtained using the HH formalism: 3860.65 MeV and 0.367 fm.

We have also reproduced the calculation of the $(L,S,I)=(0,1,0)$ $cc\bar n\bar n$ state of 
Refs.~\cite{Jan04,Sil93} using the BCN model. In Table~\ref{tres-5} (second and third columns) 
we present the energies and RMS obtained for this state for all values of $K$. For $K=26$ we have obtained an
energy of 3899.2 MeV as compared to 3904.7 MeV of Ref.~\cite{Jan04}
and 3931.0 MeV of Ref.~\cite{Sil93}. Ref.~\cite{Jan04} designed
a powerful method similar to the stochastic variational approach
to study this particular system. Although they are not fully
converged, the agreement gives confidence on both results. Ref.~\cite{Sil93}
uses a diagonalization in a restricted Hilbert space, obtaining a larger
value. 

\subsection{The $QQ \bar n \bar n$ system}
\label{sec4a}

In Ref.~\cite{Vij07} the question: Does the quark model naturally predict the
existence of $Q\bar Q n\bar n$ bound states? was posed. The answer was clear,
no compact bound four--quark state were found for any set of quantum numbers if only
two-body potentials in a complete basis were used. 
One cannot discard that
a modification of the Hilbert space, like for example considering only
diquark configurations, or of the interacting hamiltonian, like many-body
contributions, could give rise to bound states. 
Is the same conclusion still valid in the $QQ\bar n\bar n$ sector? 
To answer this question we have performed an exhaustive analysis of the 
$QQ\bar n \bar n$ spectra by means of the quark models described above. 
 Some particular results were already presented in Ref.~\cite{Vij08b},
where we studied the possibility of the $X(3872)$ being a 
$c \bar c n\bar n$ tetraquark. We gave arguments that favored the
existence of $QQ\bar n \bar n$ stable states in nature, while making
harder the existence of $Q\bar Q n\bar n$ stable states. To make the
physics clear we compared two particular sets of quantum numbers:
$J^{PC}(I)=1^{++}(0)$ for $c\bar cn\bar n$ and $J^{P}(I)=1^+(0)$ for $cc
\bar n\bar n$. Based on a variational study with a confining
mass independent many body potential we argued in Refs.~
\cite{Vij07b,Vij08b} that the binding would increase when increasing the
mass ratio of flavor exotic four-quark systems. Definitive conclusions
can only be obtained based on realistic calculations. For these purposes,
in the present work we have considered all isoscalar 
and isovector states with total orbital angular momentum $L\le 1$.
For positive (negative) parity four--quark systems the ground state corresponds to $L=0$ ($L=1$),
since parity can be expressed in terms of the relative orbital angular momenta associated 
to the Jacobi coordinates as $P=(-)^{\ell_1+\ell_2+\ell_3}$. 
This means that $P=-1$ needs three units of relative orbital
angular momentum to obtain $L=0$ ($\ell_1=\ell_2=\ell_3=1$) 
while only one is needed for $L=1$. The same reasoning applies
for $P=+1$ states. Therefore, since the complexity of the calculation and the computing 
time increase with $L$, we have not considered $L=1$ 
positive parity states that should be higher in energy.
The calculation has 
been done up to the maximum value of $K$ within our computational capabilities, $K_{\rm max}$.

In Table~\ref{tres-1} we present the $cc\bar n\bar n$ CQC
final results obtained for all possible $L=0$ isoscalar 
and isovector states and for the negative parity $L=1$ states. We indicate for each state the maximum value of $K$ used, 
$K_{\rm max}$. A first glance at this table sheds two general conclusions.
Firstly, opposite to the $Q\bar Qn\bar n$ case there exist bound states in 
the $QQ\bar n\bar n$ spectra when only two-body potentials are used. Second,
once the four--body problem 
is properly solved the number of bound states is small. Curiously, there are two 
cases that do not converge to the lowest possible two-meson threshold but to a higher one, 
the $(L=0,I=0)$ $J^{P}=0^{-}$ and $2^{-}$. These states 
get stuck in the lowest $D$ wave threshold, either the $D_1\,D_2\vert_D$ (5316 MeV) or 
the $D_J^*\,D_{J'}^*\vert_D$ (5325 MeV), satisfying $\Delta_E \ge 0$. 

As discussed in Sec. \ref{sec:compact} the convergence of unbound or
molecular states close to the two
meson threshold is a difficult numerical problem. In order to minimize this
problem we have introduced the extrapolation
formula (\ref{extra}).
In Table~\ref{tres-2}, we compare the energies obtained for $K_{\rm max}$ and 
for $K\to\infty$, using the extrapolation formula. For unbound or
loosely bound states we observe how the
extrapolation leads to four--quark energies 
within only a few MeV of the corresponding threshold.
Throughout this manuscript we shall always refer to $E_{4q}(K_{\rm max})$ as 
the four--quark state energy,~$E_{4q}$.
In particular cases we will use the extrapolation to study the characteristics
of some specific states, mainly those that are close below the threshold.

There are two ingredients that may alter the stability of the $QQ\bar n\bar n$ system:
either the mass of the heavy quark or the interacting potential. One should wonder
if the characteristic spectrum obtained in Table~\ref{tres-1} would be greatly 
influenced by them.
It was pointed out in the early 80's that a $QQ\bar Q'\bar Q'$ four--quark state should be stable against
dissociation into $Q\bar Q'+Q\bar Q'$ if the ratio $m_Q/m_{Q'}$ is large enough~\cite{Ade82}. 
This was corroborated by chiral perturbation theory 
and lattice QCD studies of the $bb\bar n\bar n$ system~\cite{Man93,Mic99}.
Trying to disentangle if a proliferation of states is predicted when the mass
of the heavy quark augments, we studied all ground states of the bottom sector
using the same interacting potential as above. The results are presented in
Table~\ref{tres-3}. We observe that all bound states become deeper than
in the charm sector and a few new states appear. Our results
strengthen the conclusion that the larger the ratio of the quark masses
the larger the binding energies.

The second ingredient to be tested is the dependence of the results on
the particular choice of the quark-quark interaction.
To answer this question we have reanalyzed all the predicted bound states 
using the BCN model. The results are summarized in
Tables~\ref{tres-3b} and~\ref{tres-3c}. The existence of bound states 
is also evident when the BCN model is
considered, although the characteristics of each state depend on the model considered.
Of particular interest is the observation that 
the bottom sector presents, independently of the quark model,
bound states with binding energies of the order of 100 MeV 
that should be possible to observe.

In connection with the interacting potential used, it has been recently
analyzed in Ref.~\cite{Vij07b} the stability of $QQ\bar n\bar n$ and
$Q\bar Q n \bar n$ systems in a simple string model considering
only a multiquark confining interaction given by the minimum
of a flip-flop or a butterfly potential in an attempt to discern whether 
confining interactions not factorizable as two--body potentials would influence 
the stability of four--quark states. The ground state of
systems made of two quarks and two antiquarks of equal masses
was found to be below the dissociation threshold. 
While for the cryptoexotic $Q\bar Q n\bar n$ the binding decreases
when increasing the mass ratio $m_Q/m_n$, for the flavor exotic $QQ\bar n\bar n$ 
the effect of mass symmetry breaking is opposite. Although more 
realistic calculations are needed before establishing a
definitive conclusion, the findings of Ref.~\cite{Vij07b} 
corroborate our results.

\subsection{Charm sector}

In the following we will study more closely those $cc\bar n\bar n$
quantum numbers, $J^P(L,S,I)$, that may host a bound state ($\Delta_E <0$
in Tables~\ref{tres-1} and~\ref{tres-3b})
trying to unveil states that might be a consequence of 
model and computational approximations.
Expected bound states are summarized in Table~\ref{tres-8}.

\subsubsection{$1^+(0,1,0)$}

The possible existence of a four--quark state with these quantum numbers was predicted more than twenty 
years ago by Zouzou and collaborators~\cite{Zou86}. Since then, several works have been devoted to 
study these particular quantum numbers by means of different methods and interactions, 
either in the charm or in the bottom sector~\cite{Car88,Pep96,Gel03,Nav07,Bri98,Jan04,Lip86}.
In Table~\ref{tres-5} we show the results obtained in this work
with the BCN and CQC models. In both cases the state converges below 
threshold. For the CQC model
the predicted binding energy is large, $-$ 76 MeV, and $\Delta_R<1$, what
would fit into the defined compact states. Opposite to that, the BCN model
predicts a rather small binding, $-$7 MeV, and the RMS is larger than one
but still not converged, although not increasing linearly. This state
would naturally correspond to a molecule.

\begin{center}
\begin{figure}
\caption{Possible electromagnetic (upper part) and weak (lower part) 
decays for $cc\bar n \bar n$ bound states. The electromagnetic
decay goes through an intermediate virtual meson (gray box) 
whose quantum numbers will depend on those of the initial 
four--quark state, virtual $D^*$ for the case of $J^P=1^+$. 
The weak decay showed corresponds to the Cabibbo allowed one 
into two kaons that incorporates two vertex 
$(c+\bar u)\to (s +\bar d)$ or $(c+\bar d)\to (\bar s +u)$. 
Other diagrams involving two $W$ emission into fermion-antifermion, 
$(cc\bar q\bar q)\to (s\bar q)+(s\bar q)+(f\bar f)+(f\bar f)$, 
or mesons, $(cc\bar q\bar q)\to (s\bar q)+(s\bar q)+(u\bar d)+(u\bar d)$, 
may also contribute \cite{Fle89}.}
\label{fig7}
\epsfig{file=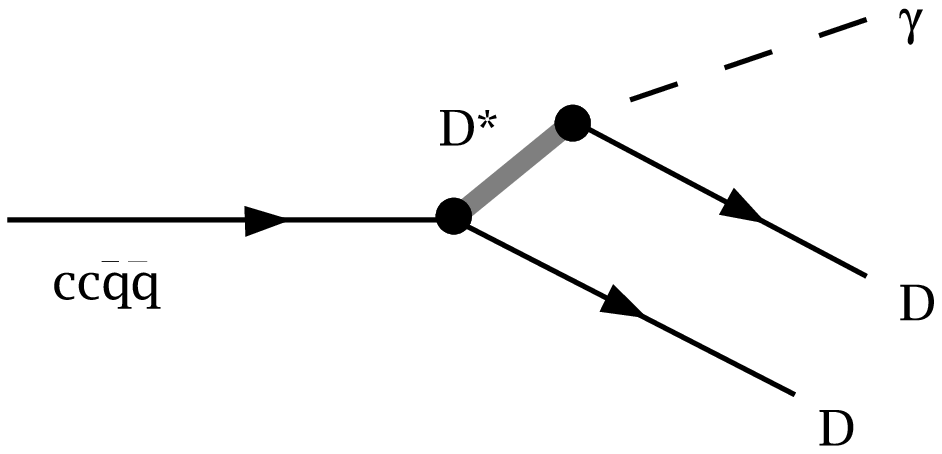, width=14cm}
\epsfig{file=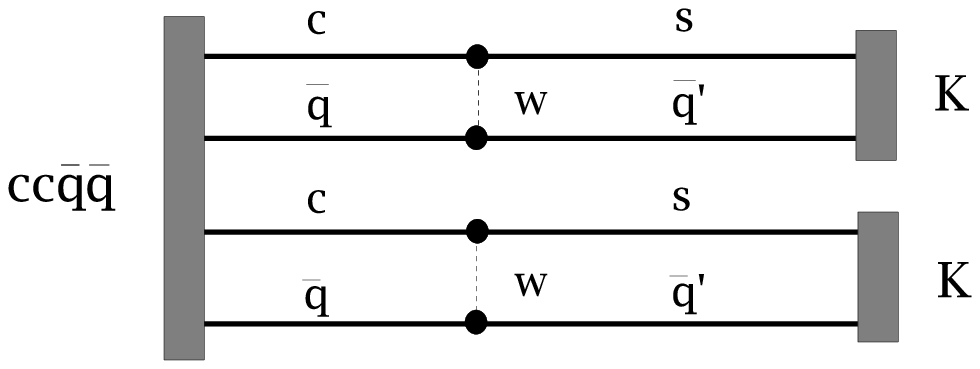, width=14cm}
\end{figure}
\end{center}

Although this state would be stable against dissociation into two mesons,
it may decay electromagnetically or weakly.
The electromagnetic transition $(QQ\bar n\bar n)\to (Q\bar n)\,(Q\bar n)\,\gamma$
would be allowed if $E_{4q}$ is 
larger than the mass of two $D$ mesons. The process is illustrated in the upper
part of Figure~\ref{fig7}. For the CQC model the 
four--quark energy is below the $D\,D$ threshold and therefore the predicted state
could only decay via a second order weak process into either 
two kaons (Cabibbo allowed) or two light mesons (Cabibbo suppressed). The process is
illustrated in the bottom part of Figure~\ref{fig7}. 
Opposite to that, the results of the BCN model allow for an 
electromagnetic decay with the emission of a photon with an energy lower than 127 MeV 
in the $cc\bar n\bar n$ rest frame. 

\subsubsection{$(1,2,3)^-(1,2,1)$
and $1^-(1,0,1)$}

Bound states have been obtained either with the CQC or the BCN models
or with both. In these cases a detailed analysis of the possible 
thresholds is required. As previously noticed, the interacting models only consider
central terms, $L$ and $S$ being proper quantum numbers.
However, as illustrated is Sec. \ref{thre} the thresholds 
may be different in the coupled or uncoupled
schemes, the former one being the relevant when trying to compare
with experiment. We show in the upper part of 
Table~\ref{tres-7} the results obtained compared to the threshold in the uncoupled
and coupled schemes. The differences are noticeable, when the coupled scheme
is used both interacting models give results above the corresponding lowest threshold,
what discard these quantum numbers as promising candidates for being observed
experimentally. 

\begin{center}
\begin{figure}
\caption{$\Delta_E$ as a function of the heavy quark 
mass, $m_Q$, for $1^-(1,0,1)$ with the CQC model. 
The grey bands take into account the differences 
between the values obtained for $K_{\rm max}$, band upper part,
and the result obtained with the extrapolation, black squares.}
\label{fig8}
\epsfig{file=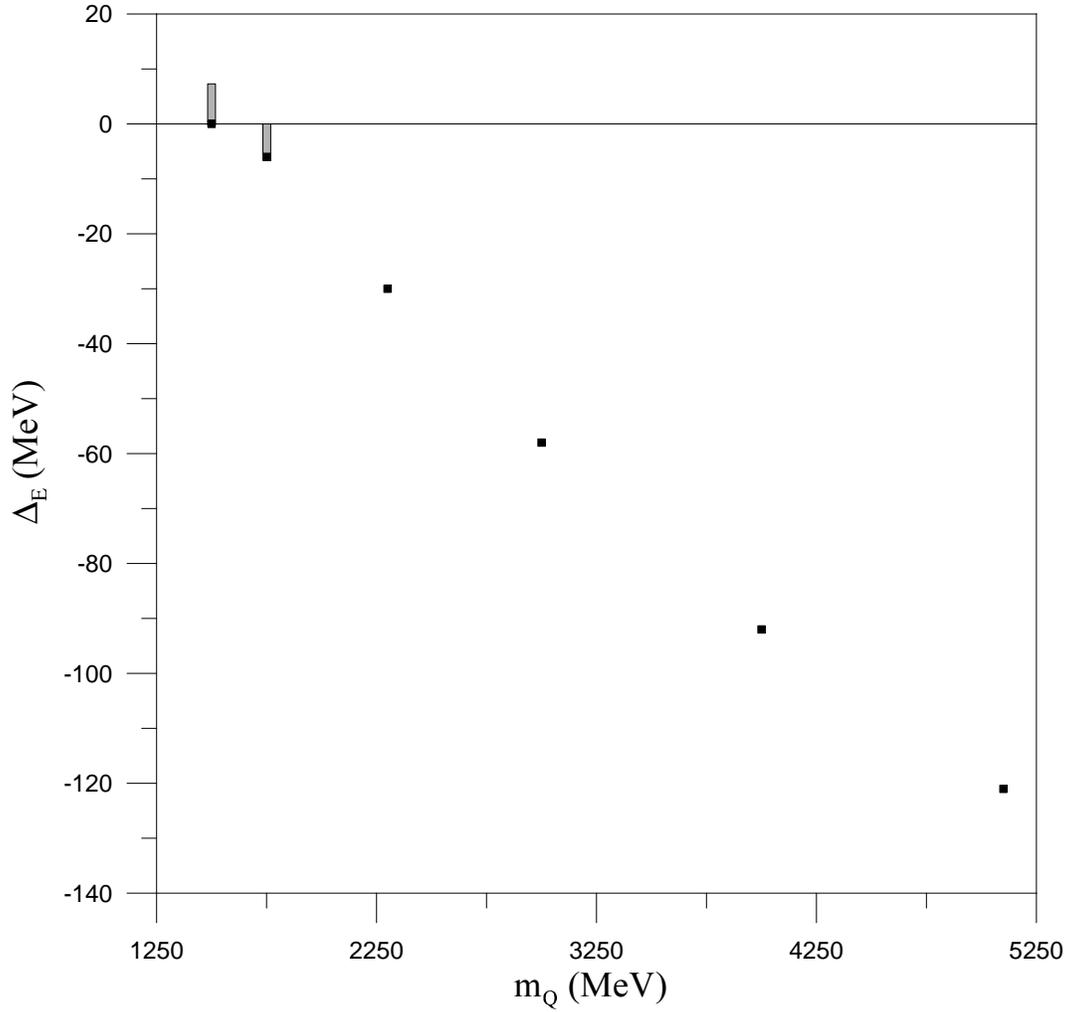,width=14cm}
\end{figure}
\end{center}

As already discussed in Sec.~\ref{sec4a},
increasing the mass of the heavy quark will favor the binding.
This is illustrated in Fig.~\ref{fig8} comparing the four-quark energy of the
$1^-(1,0,1)$ state to its uncoupled threshold in the CQC model. 
As can be seen the CQC model predicts a tiny
binding energy, but the extrapolation gives a very stable value 
around $\Delta_E\approx -6$ MeV. 
Opposite to the case of unbound states, the RMS does not grow linearly with $K$. 
We observe the changes in $\Delta_E$
when increasing $m_Q$, noting how $\Delta_E$ crosses zero for masses
slightly below the charm quark mass. As a consequence, all the states 
discussed in this section might be candidates to be stable in the 
bottom sector due to the binding gained with the heavy quark mass.

\subsection{Bottom sector}

We repeat the same analysis as before for the bottom sector.
Expected bound states are summarized in Table~\ref{tres-8}.

\subsubsection{$1^+(0,1,0)$}

Opposite to the charm sector, in the bottom sector both quark models 
offer the same prediction, a compact deeply bound four--quark state.
Within the CQC model the binding energy gets a value of $-$214 MeV while BCN predicts
a somewhat smaller value of $-$144 MeV. Concerning the possible decays, both of them are also below 
the threshold for electromagnetic decays into $B\,B$ (10588 MeV for CQC and 10602 MeV for BCN) and 
therefore the only allowed decay mode for this state will involve 
a second order weak process into two open-charm 
(Cabibbo allowed) or two light mesons (Cabibbo suppressed).

\subsubsection{$0^+(0,0,0)$ and $3^-(1,2,1)$}

The existence of a positive parity $QQ\bar n\bar n$ bound state with quantum numbers $S=0$ and 
$I=0$ was proposed in Ref.~\cite{Zou86} using the BCN model if the ratio between the masses of 
the heavy and light quarks was larger than 5$-$10, i.e., $bc\bar n\bar n$ and heavier. We have 
obtained a similar limit for the CQC model, $m_b/m_n\approx15$. 

For the $J^P=3^-$ state a bound state has also been found using both models, being
$\Delta_E=-140$ MeV and $\Delta_E=-119$ MeV, respectively. 
As can be seen in Table~\ref{tres-7} it is the only member of a multiplet 
below all possible thresholds and therefore all strong decays are forbidden. 

No bound states were observed in the charm sector with these quantum numbers,
being therefore consequence of the binding gained due to the larger heavy quark
mass. Both states would present electromagnetic decays, the former one
$(bb\bar n\bar n)\,\,\to (B\,B\vert_P)\,\,\gamma$, 
with a photon energy in the range of $\lesssim 400$ MeV,
and the latter through $(bb\bar n\bar n) \to (B\,B\vert_D)\,\,\gamma$ or 
$(bb\bar n\bar n)\,\,\to (B^*\,B^*\vert_S)\,\,\gamma$ 
emitting a photon of the order of $\lesssim 400-450$ MeV.

\subsubsection{$1^-(1,0,0)$}

Although no bound state is observed in the $cc\bar n\bar n$ spectra, 
a bound state, $\Delta_E=-11$ MeV, appears in the bottom sector with the CQC model.
The structure of the wave function, $\Delta_R\approx1.182$, clearly 
points to an extended meson--meson molecule instead of a compact four--quark state. 
Concerning its possible decay channels, being this state below all possible thresholds it could 
only undergo a weak decay.

\subsubsection{$1^-(1,0,1)$, $2^+(0,2,1)$ and $(1,2)^-$ $(1,2,1)$}

All these states are predicted to be bound in the uncoupled scheme but none
of them survive the coupled thresholds. Although interesting from the theoretical point 
of view to carefully test our calculating framework, they are not expected to be observed in nature.

\subsection{Beyond $QQ\bar n\bar n$}
Once the $QQ\bar n\bar n$ ($Q=c,b$) states have been discussed, few avenues remain to be explored. Among
them, the consideration of $QQ\bar s\bar s$ and $QQ'\bar n\bar n$ states would be of interest. 
Concerning the strangeness $\pm2$ states only those containing
$b$ quarks seem likely candidates to accommodate bound states. 
We have redone the calculation for the most promising states,
those with quantum numbers $I(J^P)=0(1^+)$ and $Q=c$ or $b$. In both
cases the four-quark system is above the corresponding threshold. 
This can be easily understood because the mass ratio $M_Q/m_q$ has diminished
due to the large mass of the strange quark, thus increasing the contribution 
of the kinetic energy. Besides, the attractive one-pion exchange does not contribute.
Our conclusions coincide with Ref.~\cite{Ebe07},
the binding energy decreases when decreasing the mass ratio $M_Q/m_q$.
The second system, $QQ'\bar n\bar n$ is made of distinguishable heavy quarks, 
i.e., $bc\bar n\bar n$ among others, and therefore demands a modification 
of the current formalism, namely consider all possible combinations of $\ell_1,S_{12}$.
This would lead to an increase in the basis size and consequently in
computing time. However, one can draw a simple conclusion in light
of Ref.~\cite{Ebe07}, or the first six lines
of Table III. The binding energy increases with the reduced mass of
the heavy quark pair. Thus, one would expect a bound state with quantum numbers
$1^+(0,1,0)$ for the $bc\bar n\bar n$ system, while the others 
appearing in the doubly-bottom sector would not be clear candidates
for bound states with a $bc$ heavy diquark.

\subsection{Meson-meson probabilities}

We present in Table~\ref{pro} the meson-meson probabilities for
some selected four--quark states according to Eqs. (\ref{proeq}). 
These calculations where done by means of the variational method, Ref.~\cite{Vij05b}.
As can be seen there is a perfect agreement with the HH results but the variational formalism
allows to evaluate the probabilities of the different physical components in a simpler manner. 
Unbound states converge to two isolated mesons, the lowest threshold of 
the system, its RMS and $\Delta_R$ being 
very large. In contrast, bound states have a radius smaller than the threshold and they 
present probabilities different from zero for several physical states, the lowest two-meson
threshold being contained in the physical four--quark system. Such states would be called
compact in our notation. When the binding energy approaches the threshold, the probability of 
a single physical channel converges to one, what we defined as a molecular state.

\subsection{BCN vs. CQC}
\label{ogeope}

The delicate interplay between the OGE and the chiral interactions in the description of 
the hadron spectra and baryon-baryon interaction has been widely discussed in the 
literature~\cite{Nak98}. In the $QQ\bar n\bar n$ spectra the relative strength of these 
interactions is even more important since $\Delta_E$ is very sensitive to any modification 
in the chiral/OGE rate. The reason is that the threshold is not affected by the chiral
part of the interaction since it is made of two $Q\bar n$ mesons, 
and there are no boson exchanges between heavy and light quarks.
This is not the case for the $QQ\bar n\bar n$ systems, 
where bosons may be exchanged between the two light antiquarks. Therefore, 
an arbitrary modification in the strength of the chiral part of the
interaction in any four--quark state where it is attractive,
could bind the system. 

We therefore emphasize the importance of testing any model against as many observables 
as possible in order to constraint its parameters and ingredients. In this respect the heavy--light 
four--quark states are ideally suited for this task. Since only the total energy, and not the 
threshold, depends on the boson
exchanges, the comparison between the predicted and measured $\Delta_E$ 
would provide us with precise information regarding the role played by these interactions in 
the meson spectra.

\begin{center}
\begin{figure}
\caption{$\Delta_R$ as a function of $K$ for the 
$2^+(0,2,1)$ $bb\bar n\bar n$ with 
boson exchange potentials (solid line) and without (dashed line) for the CQC model.}
\label{fig3}
\vspace*{0.7cm}
\hspace*{-0.7cm}
\epsfig{file=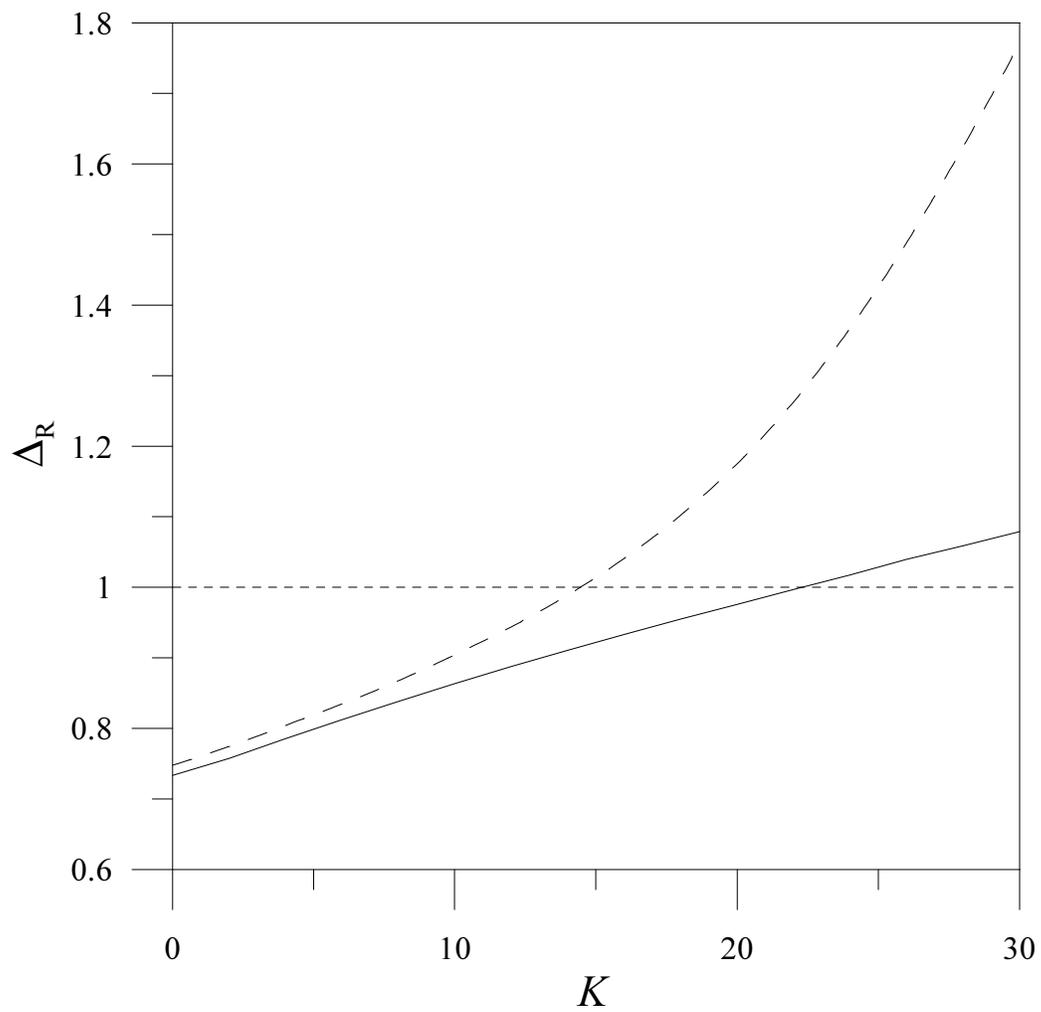, width=14cm}
\end{figure}
\end{center}

Amazingly, as observed in Table~\ref{tres-8} almost all bound states are predicted independently
of the dynamics. In other words, whenever a bound state is found with the CQC model,
the BCN potential predicts a similar state. In general predicted 
binding energies are smaller for gluon-based interactions. Only
the $1^-(1,0,0)$ $bb\bar n\bar n$ state is predicted by one of the models, the CQC.
To illustrate the larger binding predicted by models considering boson exchanges
we have selected one of the states whose binding energy 
in the uncoupled scheme has been found 
to be smaller than 10 MeV, the $bb\bar n\bar n$ $2^+(0,2,1)$. 
Once boson exchanges are switched off the state lacks of enough attraction to be bound, 
behaving like two isolated mesons.
These results are illustrated in Table \ref{tres-9}.
To illustrate the difference in the structure induced by the 
boson exchanges we have plotted in Fig.~\ref{fig3} the 
evolution with $K$ of $\Delta_R$ in both cases. 
In the unbound case the system is separating very rapidly while the bound one starts 
to converge to a large, but finite, value.

\section{Experimental Observation}
\label{Experiment}

The most promising mechanism for the production of a four--quark state is the independent
formation of two $cc$ and $\bar q\bar q$ pairs to later on merge into a $cc\bar q\bar q$ state. 
Following this idea the rate of $cc\bar q\bar q$ production was estimated in Ref.~\cite{Gel03} as
$R_{cc\bar q\bar q}/R_{ccq}\approx 1/10$. Therefore, any facility able to produce double charm baryons in 
sizable quantities should be able, in principle, to observe four--quark states. 

So far, double charmed baryons have only been observed by SELEX Collaboration \cite{Sel02}, although its 
limited statistics, $\approx 50$ events, makes doubtful that four--quark states can be produced in quantities
large enough to be statistically significant. In Ref.~\cite{Sch03} the production rate of double charmed baryons by 
the COMPASS experiment was estimated to be of the order of $10^4$ to 
$1.7\cdot 10^4$ events, this will indicate that
up to 1500 four--quark events may be produced by COMPASS 200 GeV proton beam~\cite{Moi08}. 
The number of $ccq$ and $bbq$ events at Tevatron was estimated in Ref.~\cite{Ani02}, obtaining values of 
10$^5$ and 10$^4$ events respectively. This would yield values 
of the order of $10^4$ four--quark events to be produced in the second run of Tevatron. Although so far Belle 
Collaboration has not reported the observation of double charmed baryons, their production cross section was estimated 
in Ref.~\cite{Ber04}, where values of the order of $10^4$ events/year were obtained. This will translate 
into 1000 four--quark events/year. Theoretical predictions about the production cross-section of double 
charmed baryons at BaBar has been estimated, giving diverse values from hundreds to tens of thousands of 
events~\cite{Edw07}. This will indicate that BaBar could produce up to 5000 four--quark/events. On 
this respect it is worth to mention that theoretical cross-section predictions for double $c\bar c$ production 
have been found to be one order of magnitude too low as compared with experimental data~\cite{Aub05}. This will 
imply that the predicted rate for double charmed baryon production may have been underestimated and therefore 
the expected number of four--quark events could be larger. In Ref.~\cite{Fab06}, the $cc\bar q\bar q$ production 
rate at different facilities was estimated assuming that the dominant mechanism for double charm production at high 
energy colliders would be a disconnected double gluon-gluon fusion, $(g+g)+(g+g)\to (c+\bar c)+(c+\bar c)$. The 
predicted number of four--quark events produced at LHC by either LHCb or ALICE and at Tevatron is found to be 
very large, 9700, 20900, and 600 events/hour respectively, while for RHIC they expect a smaller value of 12 events/hour.

The picture that seems to emerge from these estimations is that nowadays we are on the verge where actual 
experimental facilities may be capable of start disentangling the properties of $cc\bar q\bar q$ bound 
states. If this is not so, the future facilities that will be in operation in the next decade, if not sooner, 
at CERN and Fermilab will be able to provide a definitive answer to the existence of 
four--quark flavor--exotic states in nature.

\section{Summary}
\label{summary}

In this work we have performed a systematic analysis of all $cc\bar n\bar n$ and $bb\bar n\bar n$ ground
states within the framework of the hyperspherical harmonic method. In order to distinguish between 
unbound, compact, and molecular four--quark states we have considered two different quark models
widely used in the literature. We have analyzed both isoscalar and isovector systems with $S$=0, 1, and 2. 
We have considered $L=0$ states with positive and negative parity and $L=1$ negative parity ones.
The relevance of a careful analysis of the numerical thresholds together with 
the numerical approximations involved has been emphasized in order to avoid 
the misidentification of bound states. Estimations about the possibility 
to detect these states on the next generation 
of experimental facilities have been performed.

Our results are summarized in Table~\ref{tres-8}. We have found five four-quark states that
should be narrow and therefore possible to detect. Four of them are predicted independently
of the interacting potential used, either CQC or BCN. The $1^-(1,0,0)$ state is
found only with the CQC model. The $cc\bar n\bar n$ system would only have one bound
state while up to four could be stable in the $bb\bar n\bar n$ system.
All predicted states are compact, only the $1^-(1,0,0)$ being molecular.
Unfortunately, only one of them, the $cc\bar n\bar n$ state with quantum 
numbers $1^+(0,1,0)$, is within the scope of the experimental 
facilities that will be available in the near future.

Theoretical models point out to the existence 
of a double charmed isoscalar four--quark bound state with 
quantum numbers $J^P=1^+$, its properties depending on 
the quark model considered. 
The experimental detection and analysis of four--quark 
double charmed states will undoubtedly prove to be an 
invaluable testing ground to severely constraint the 
different theoretical models and 
therefore, it will allow to refine the theoretical 
predictions all over the hadronic spectra.

\section{Acknowledgments}
JV thanks M. Moinester for stimulating discussions regarding double charm production.
This work has been partially funded by the Spanish Ministerio de
Educaci\'on y Ciencia and EU FEDER under Contract No. FPA2007-65748,
by Junta de Castilla y Le\'{o}n under Contracts No. SA016A17 and Grupos
de Excelencia GR12, 
and by the Spanish Consolider-Ingenio 2010 Program CPAN (CSD2007-00042).

\begin{center}
\begin{table}[ht]
\caption{Energy difference, in MeV, between four--quark states $QQ\bar n\bar n$ ($Q=c$ or $b$) 
and the corresponding two-meson threshold for different theoretical approaches.}
\label{tres-10}
\begin{tabular}{|c|ccc|ccc||ccc|ccc|}
\hline
        &\multicolumn{6}{|c||}{$cc\bar n\bar n$}&\multicolumn{6}{|c|}{$bb\bar n\bar n$}\\
\hline
        &\multicolumn{3}{|c|}{$I=0$}&\multicolumn{3}{|c||}{$I=1$}&\multicolumn{3}{|c|}{$I=0$}&\multicolumn{3}{|c|}{$I=1$}\\
\hline
$J^P$           & $0^+$ & $1^+$ & $2^+$ & $0^+$ & $1^+$ & $2^+$ &$0^+$  & $1^+$ & $2^+$ &$0^+$  & $1^+$ & $2^+$ \\
\hline
\cite{Zou86}    &       & $>0$  &       &       &       &       &       & $-106$&       &       &       &       \\
\cite{Lip86}    &       &$\lesssim0$&   &       &       &       &       & $<0$  &       &       &       &       \\
\cite{Car88}    &       &$>0$   &       &       &       &       &       & $\approx-70$  &       &       &       &       \\
\cite{Man93}    &       &       &       &       &       &       &       &$-8.3$ &       &       &       &       \\
\cite{Sil93}    &$>+60$ &+19    &$>+60$ &$>+60$ &$>+60$ &$>+60$ &$>+60$ & $-131$&$>+60$ &$>+60$ &$+56$  &$+30$  \\
\cite{Pep96}\footnotemark\footnotetext{Results for two different set of
        parameters $C_1$ ($C_2$).}      
                &       &$-$185 ($-$332)        &       &       &       &       &       & $-226$ ($-$497) &     &       &       &       \\
\cite{Bri98}    &       &       &       &       &       &       &       & $-99$ &       &+156   &+117   &+86    \\
\cite{Gel03}    &       &$-(30-60)$&&   &       &       &       &       &       &       &       &       \\
\cite{Jan04}\footnotemark\footnotetext{Results for two different interacting potentials.}
                &       &$-0.6$ ($-2.7$)        &       &       &       &       &       &$-132$ ($-$140)        &       &       &       &       \\
\cite{Vij04b}   &+585   &$-129$ &+830   &+384   &+293   &+192   &+258   &$-341$ &+708   &+128   &+96    &+65 \\
\cite{Nav07}    &       &$+125\pm200$&  &       &       &       &       &$-400\pm300$&  &       &       &       \\
\cite{Ebe07}    &       &+64    &       &+327   &+208   &+104   &       &$-102$ &       &+90    &+53    &+23    \\
\cite{Zha07}\footnotemark\footnotetext{Results for two different interacting potentials.}
                &       &+50 (+60)&      &+143   &+299   &+213   &       &$-32$ $(-18)$&    &+119   &+93    &+72\\
\hline
\end{tabular}
\end{table}
\end{center}

\begin{center}
\begin{table}[ht!!]
\caption{$D$ and $B$ meson masses (in MeV) and root mean square radius, RMS, (in fm) obtained with 
the quark models described in Sec. \ref{tech}. Experimental masses (Exp.) 
are taken from Ref.~\protect\cite{PDG06}, except for the state denoted by a dagger
that has been taken from Ref.~\protect\cite{Belb4}.}
\label{tmeson}
\begin{tabular}{|cc|c|cc|cc|cc|}
\hline
                & &     Exp.    & \multicolumn{2}{|c|}{CQC$_{18}$}& \multicolumn{2}{|c|}{CQC} & \multicolumn{2}{|c|}{BCN}\\
($L,S,J$)&State         &                       & E     & RMS   & E     & RMS   & E     & RMS   \\
\hline
(0,0,0) &$D$            & 1864.5$\pm$0.4        & 1883          & 0.207         & 1936  & 0.220 & 1886  & 0.212\\
(0,1,1) &$D^*$          & 2006.7$\pm$0.4        & 2010          & 0.237         & 2001  & 0.234 & 2020  & 0.235\\
(1,1,0) &$D^*_0$        & 2308.0$\pm$17$\pm$12$^{\dagger}$      & 2465  & 0.344 & 2498  & 0.373 & 2491  & 0.342\\
(1,0,1) &$D_1$          & 2422.3$\pm$1.3        & 2492          & 0.370         & 2490  & 0.369 & 2455  & 0.332\\
(1,1,1) &$D^*_1$        & 2427$\pm$40           & 2504          & 0.368         & 2498  & 0.373 & 2491  & 0.342\\
(1,1,2) &$D^*_2$        & 2461.1$\pm$1.6        & 2496          & 0.380         & 2498  & 0.373 & 2491  & 0.342\\
\hline
(0,0,0) &$B$            & 5279.0$\pm$0.5        & 5281          & 0.139         & 5294  & 0.142 & 5301  & 0.141\\
(0,1,1) &$B^*$          & 5325.0$\pm$0.6        & 5321          & 0.146         & 5318  & 0.145 & 5350  & 0.147\\
(1,1,0) &$B^*_0$        & 5698$\pm$8
\footnotemark\footnotetext{According to the Particle Data book~\cite{PDG06} this signal can be interpreted as stemming 
from several narrow and broad resonances in the range 5650$-$5750 MeV. No quantum numbers are given.}
                & 5848          & 0.230         & 5810  & 0.232         &  5825 & 0.217\\
(1,0,1) &$B_1$          & 5698$\pm$8            & 5768          & 0.239         & 5807  & 0.231 & 5811  & 0.214\\
(1,1,1) &$B^*_1$        & 5698$\pm$8            & 5876          & 0.232         & 5810  & 0.232 & 5825  & 0.217\\
(1,1,2) &$B^*_2$        & 5698$\pm$8            & 5786          & 0.231         & 5810  & 0.232 & 5825  & 0.217 \\
\hline
\end{tabular}
\end{table}
\end{center}

\begin{center}
\begin{table}
\caption{Experimental lowest two-meson thresholds for charmed and bottom four--quark states 
in the uncoupled (UN) and coupled (CO) schemes as defined in the text. 
$M_1\,M_2\vert_L$ indicates the lowest threshold and $E(M_1,M_2)$ its energy.
Energies are in MeV.}
\label{th1}
\begin{tabular}{|c|cc|cc||cc|cc|}
\hline
                &\multicolumn{4}{c||}{$cc\bar n\bar n$} &\multicolumn{4}{c|}{$bb\bar n\bar n$}\\
\hline
                &\multicolumn{2}{c|}{UN}        &\multicolumn{2}{c||}{CO} 
                &\multicolumn{2}{c|}{UN}        &\multicolumn{2}{c|}{CO} \\
\hline
$J^P(I)$        & $M_1\,M_2\vert_L$                     & $E(M_1,M_2)$  &$M_1\,M_2\vert_L$ &$E(M_1,M_2)$
                & $M_1\,M_2\vert_L$                     & $E(M_1,M_2)$  &$M_1\,M_2\vert_L$ &$E(M_1,M_2)$\\ 
\hline
\multicolumn{9}{|c|}{$L=0$}\\
\hline
$0^+(0)$        & $D_1\,D\vert_P$               & 4290                  & $D_1\,D\vert_P$       & 4290 
                & $B_1\,B\vert_P$               & 10977                 & $B_1\,B\vert_P$       & 10977 \\
$0^+(1)$        & $D\,D\vert_S$                 & 3735                  & $D\,D\vert_S$         & 3735 
                & $B\,B\vert_S$                 & 10558                 & $B\,B\vert_S$         & 10588 \\ 
$1^+(0,1)$      & $D\,D^*\vert_S$               & 3877                  & $D\,D^*\vert_{S,D}$   & 3877 
                & $B\,B^*\vert_S$               & 10604                 & $B\,B^*\vert_{S,D}$   & 10604 \\ 
$2^+(0)$        & $D^*\,D_0^*\vert_P$           & 4317                  & $D\,D^*\vert_D$       & 3877 
                & $B^*\,B_0^*\vert_P$           & 11023                 & $B\,B^*\vert_D$       & 10604 \\
$2^+(1)$        & $D^*\,D^*\vert_S$             & 4018                  & $D\,D\vert_D$         & 3735 
                & $B^*\,B^*\vert_S$             & 10650                 & $B\,B\vert_D$         & 10558 \\ 
$0^-(0,1)$      & $D_0^*\,D_1^*\vert_P$         & 4735                  & $D\,D^*\vert_P$       & 3877 
                & $B_0^*\,B_1^*\vert_P$         & 11396                 & $B\,B^*\vert_P$       & 10604 \\ 
$1^-(0)$        & $D_1\,D_0^*\vert_P$           & 4730                  & $D\,D\vert_P$         & 3735 
                & $B_1\,B_0^*\vert_P$           & 11396                 & $B\,B\vert_P$         & 10558 \\ 
$1^-(1)$        & $D_0^*\,D_0^*\vert_P$         & 4616                  & $D\,D^*\vert_P$       & 3877 
                & $B_0^*\,B_0^*\vert_P$         & 11396                 & $B\,B^*\vert_P$       & 10604 \\
$2^-(0,1)$      & $D_0^*\,D_1^*\vert_P$         & 4735                  & $D\,D^*\vert_P$       & 3877 
                & $B_0^*\,B_1^*\vert_P$         & 11396                 & $B\,B^*\vert_P$       & 10604 \\
\hline
\multicolumn{9}{|c|}{$L=1$ $S=0$}\\
\hline
$1^-(0)$        & $D\,D\vert_P$                 & 3735                  & $D\,D\vert_P$         & 3735 
                & $B\,B\vert_P$                 & 10558                 & $B\,B\vert_P$         & 10558 \\
$1^-(1)$        & $D\,D_1\vert_{S,D}$           & 4290                  & $D\,D^*\vert_P$       & 3877 
                & $B\,B_1\vert_{S,D}$           & 10977                 & $B\,B^*\vert_P$       & 10604 \\
\hline
\multicolumn{9}{|c|}{$L=1$ $S=1$}\\
\hline
$0^-(0,1)$      & $D\,D^*\vert_P$               & 3877                  & $D\,D^*\vert_P$       & 3877 
                & $B\,B^*\vert_P$               & 10604                 & $B\,B^*\vert_P$       & 10604 \\ 
$1^-(0)$        & $D\,D^*\vert_P$               & 3877                  & $D\,D\vert_P$         & 3735  
                & $B\,B^*\vert_P$               & 10604                 & $B\,B\vert_P$         & 10558 \\ 
$1^-(1)$        & $D\,D^*\vert_P$               & 3877                  & $D\,D^*\vert_P$       & 3877  
                & $B\,B^*\vert_P$               & 10604                 & $B\,B^*\vert_P$       & 10604 \\ 
$2^-(0,1)$      & $D\,D^*\vert_P$               & 3877                  & $D\,D^*\vert_P$       & 3877  
                & $B\,B^*\vert_P$               & 10604                 & $B\,B^*\vert_P$       & 10604 \\ 
\hline
\multicolumn{9}{|c|}{$L=1$ $S=2$}\\
\hline
$1^-(0)$        & $D^*\,D^*\vert_P$             & 4018                  & $D\,D\vert_P$         & 3735 
                & $B^*\,B^*\vert_P$             & 10650                 & $B\,B\vert_P$         & 10558 \\
$1^-(1)$        & $D^*\,D_0^*\vert_{S,D}$       & 4317                  & $D\,D^*\vert_P$       & 3877  
                & $B^*\,B_0^*\vert_{S,D}$       & 11023                 & $B\,B^*\vert_P$       & 10604 \\
$2^-(0)$        & $D^*\,D^*\vert_P$             & 4018                  & $D\,D^*\vert_P$       & 3877 
                & $B^*\,B^*\vert_P$             & 10650                 & $B\,B^*\vert_P$       & 10604 \\
$2^-(1)$        & $D^*\,D_0^*\vert_D$           & 4317                  & $D\,D^*\vert_P$       & 3877 
                & $B^*\,B_0^*\vert_D$           & 11023                 & $B\,B^*\vert_P$       & 10604 \\
$3^-(0)$        & $D^*\,D^*\vert_P$             & 4018                  & $D^*\,D^*\vert_P$     & 4018 
                & $B^*\,B^*\vert_P$             & 10650                 & $B^*\,B^*\vert_P$     & 10650 \\
$3^-(1)$        & $D^*\,D_0^*\vert_D$           & 4317                  & $D\,D_1\vert_D$       & 4290  
                & $B^*\,B_0^*\vert_D$           & 11023                 & $B\,B_1\vert_D$       & 10977 \\
\hline
\end{tabular}
\end{table}
\end{center}

\begin{center}
\begin{table}
\caption{Same as Table \ref{th1} for $cc\bar n\bar n$ states with the CQC model. 
We have evaluated the RMS, in fm.}
\label{th2}
\begin{tabular}{|c|ccc|ccc|}
\hline
                &\multicolumn{3}{c|}{UN}        &\multicolumn{3}{c|}{CO} \\
\hline
$J^P(I)$        & $M_1\,M_2\vert_L$     & $E(M_1,M_2)$&RMS      &$M_1\,M_2\vert_L$      &$E(M_1,M_2)$&RMS\\ 
\hline
\multicolumn{7}{|c|}{$L=0$}\\
\hline
$0^+(0)$        & $D_1\,D\vert_P$       & 4426                  & 0.589 & $D_1\,D\vert_P$       & 4426          & 0.589 \\
$0^+(1)$        & $D\,D\vert_S$         & 3872                  & 0.440 & $D\,D\vert_S$         & 3872          & 0.440 \\ 
$1^+(0,1)$      & $D\,D^*\vert_S$       & 3937                  & 0.454 & $D\,D^*\vert_{S,D}$   & 3937          & 0.454 \\ 
$2^+(0)$        & $D^*\,D_J^*\vert_P$   & 4499                  & 0.607 & $D\,D^*\vert_D$       & 3937          & 0.454 \\
$2^+(1)$        & $D^*\,D^*\vert_S$     & 4002                  & 0.468 & $D\,D\vert_D$         & 3872          & 0.440 \\ 
$0^-(0)$        & $D_1\,D_1\vert_P$     & 4980                  & 0.738 & $D\,D^*\vert_P$       & 3937          & 0.454 \\ 
$0^-(1)$        & $D_J^*\,D_J^*\vert_P$ & 4996                  & 0.746 & $D\,D^*\vert_P$       & 3937          & 0.454 \\ 
$1^-(0)$        & $D_1\,D_J^*\vert_P$   & 4988                  & 0.742 & $D\,D\vert_P$         & 3872          & 0.440 \\ 
$1^-(1)$        & $D_1\,D_J^*\vert_P$   & 4988                  & 0.742 & $D\,D^*\vert_P$       & 3937          & 0.454 \\ 
$2^-(0,1)$      & $D_J^*\,D_J^*\vert_P$ & 4996                  & 0.746 & $D\,D^*\vert_P$       & 3937          & 0.454 \\
\hline
\multicolumn{7}{|c|}{$L=1$ $S=0$}\\
\hline
$1^-(0)$        & $D\,D\vert_P$         & 3872                  & 0.440 & $D\,D\vert_P$         & 3872          & 0.440 \\
$1^-(1)$        & $D\,D_1\vert_{S,D}$   & 4426                  & 0.589 & $D\,D^*\vert_P$       & 3937          & 0.454 \\
\hline
\multicolumn{7}{|c|}{$L=1$ $S=1$}\\
\hline
$0^-(0,1)$      & $DD^*\vert_P$         & 3937                  & 0.454 & $D\,D^*\vert_P$       & 3937          & 0.454 \\
$1^-(0)$        & $D\,D^*\vert_P$       & 3937                  & 0.454 & $D\,D\vert_P$         & 3872          & 0.440 \\
$1^-(1)$        & $D\,D^*\vert_P$       & 3937                  & 0.454 & $D\,D^*\vert_P$       & 3937          & 0.454 \\
$2^-(0,1)$      & $D\,D^*\vert_P$       & 3937                  & 0.454 & $D\,D^*\vert_P$       & 3937          & 0.454 \\ 
\hline
\multicolumn{7}{|c|}{$L=1$ $S=2$}\\
\hline
$1^-(0)$        & $D^*\,D^*\vert_P$     & 4002                  & 0.468 & $D\,D\vert_P$         & 3872          & 0.440\\
$1^-(1)$        & $D^*\,D_J^*\vert_{S,D}$& 4499                 & 0.607 & $D\,D^*\vert_P$       & 3937          & 0.454\\ 
$2^-(0)$        & $D^*\,D^*\vert_P$     & 4002                  & 0.468 & $D\,D^*\vert_P$       & 3937          & 0.454\\
$2^-(1)$        & $D^*\,D_J^*\vert_{S,D}$& 4499                 & 0.607 & $D\,D^*\vert_P$       & 3937          & 0.454\\
$3^-(0)$        & $D^*\,D^*\vert_P$     & 4002                  & 0.468 & $D^*\,D^*\vert_P$     & 4002          & 0.468 \\
$3^-(1)$        & $D^*\,D_J^*\vert_{S,D}$& 4499                 & 0.607 & $D_1\,D\vert_D$       & 4426          & 0.589\\ 
\hline
\end{tabular}
\end{table}
\end{center}

\begin{center}
\begin{table}
\caption{Same as Table \ref{th2} for $bb \bar n\bar n$ states.}
\label{th3}
\begin{tabular}{|c|ccc|ccc|}
\hline
                &\multicolumn{3}{c|}{UN}        &\multicolumn{3}{c|}{CO} \\
\hline
$J^P(I)$        & $M_1\,M_2\vert_L$     & $E(M_1,M_2)$ &RMS     &$M_1\,M_2\vert_L$      &$E(M_1,M_2)$ &RMS\\ 
\hline
\multicolumn{7}{|c|}{$L=0$}\\
\hline
$0^+(0)$        & $B_1\,B\vert_P$       & 11101         & 0.373 & $B_1\,B\vert_P$       & 11101         & 0.373 \\
$0^+(1)$        & $B\,B\vert_S$         & 10588         & 0.284 & $B\,B\vert_S$         & 10588         & 0.284 \\ 
$1^+(0,1)$      & $B\,B^*\vert_S$       & 10612         & 0.287 & $B\,B^*\vert_{S,D}$   & 10612          & 0.287        \\ 
$2^+(0)$        & $B^*\,B_J^*\vert_P$   & 11128         & 0.377 & $B\,B^*\vert_D$       & 10612         & 0.287 \\
$2^+(1)$        & $B^*\,B^*\vert_S$     & 10636         & 0.291 & $B\,B\vert_D$         & 10588         & 0.284 \\ 
$0^-(0)$        & $B_1\,B_1\vert_P$     & 11614         & 0.462 & $B\,B^*\vert_P$       & 10612         & 0.287 \\ 
$0^-(1)$        & $B_J^*\,B_J^*\vert_P$ & 11620         & 0.464 & $B\,B^*\vert_P$       & 10612         & 0.287 \\ 
$1^-(0)$        & $B_1\,B_J^*\vert_P$   & 11617         & 0.463 & $B\,B\vert_P$         & 10588         & 0.284 \\ 
$1^-(1)$        & $B_1\,B_J^*\vert_P$   & 11617         & 0.463 & $B\,B^*\vert_P$       & 10612         & 0.287 \\ 
$2^-(0,1)$      & $B_J^*\,B_J^*\vert_P$ & 11620         & 0.464 & $B\,B^*\vert_P$       & 10612         & 0.287 \\
\hline
\multicolumn{7}{|c|}{$L=1$ $S=0$}\\
\hline
$1^-(0)$        & $B\,B\vert_P$         & 10588         & 0.284 & $B\,B\vert_P$         & 10588         & 0.284 \\
$1^-(1)$        & $B\,B_1\vert_{S,D}$   & 11101         & 0.373 & $B\,B^*\vert_P$       & 10612         & 0.287 \\
\hline
\multicolumn{7}{|c|}{$L=1$ $S=1$}\\
\hline
$0^-(0,1)$      & $BB^*\vert_P$         & 10612         & 0.287 & $B\,B^*\vert_P$       & 10612         & 0.287 \\
$1^-(0)$        & $B\,B^*\vert_P$       & 10612         & 0.287 & $B\,B\vert_P$         & 10588         & 0.284 \\
$1^-(1)$        & $B\,B^*\vert_P$       & 10612         & 0.287 & $B\,B^*\vert_P$       & 10612         & 0.287 \\
$2^-(0,1)$      & $B\,B^*\vert_P$       & 10612         & 0.287 & $B\,B^*\vert_P$       & 10612         & 0.287 \\ 
\hline
\multicolumn{7}{|c|}{$L=1$ $S=2$}\\
\hline
$1^-(0)$        & $B^*\,B^*\vert_P$     & 10636         & 0.291 & $B\,B\vert_P$         & 10588         & 0.284\\
$1^-(1)$        & $B^*\,B_J^*\vert_{S,D}$& 11128        & 0.377 & $B\,B^*\vert_P$       & 10612         & 0.287\\ 
$2^-(0)$        & $B^*\,B^*\vert_P$     & 10636         & 0.291 & $B\,B^*\vert_P$       & 10612         & 0.287\\
$2^-(1)$        & $B^*\,B_J^*\vert_{S,D}$& 11128        & 0.377 & $B\,B^*\vert_P$       & 10612         & 0.287\\
$3^-(0)$        & $B^*\,B^*\vert_P$     & 10636         & 0.291 & $B^*\,B^*\vert_P$     & 10636         & 0.291 \\
$3^-(1)$        & $B^*\,B_J^*\vert_{S,D}$& 11128        & 0.377 & $B_1\,B\vert_D$       & 11101         & 0.373\\ 
\hline
\end{tabular}
\end{table}
\end{center}

\begin{center}
\begin{table}
\caption{Same as Table \ref{th2} for the BCN model.}
\label{th2b}
\begin{tabular}{|c|ccc|ccc|}
\hline
                &\multicolumn{3}{c|}{UN}        &\multicolumn{3}{c|}{CO} \\
\hline
$J^P(I)$        & $M_1\,M_2\vert_L$     & $E(M_1,M_2)$ &RMS     &$M_1\,M_2\vert_L$      & $E(M_1,M_2)$ &RMS\\ 
\hline
\multicolumn{7}{|c|}{$L=0$}\\
\hline
$0^+(0)$        & $D_1\,D\vert_P$       & 4341          & 0.544 & $D_1\,D\vert_P$       & 4341          & 0.544 \\
$0^+(1)$        & $D\,D\vert_S$         & 3772          & 0.424 & $D\,D\vert_S$         & 3772          & 0.424 \\ 
$1^+(0,1)$      & $D\,D^*\vert_S$       & 3906          & 0.447 & $D\,D^*\vert_{S,D}$   & 3906          & 0.447 \\ 
$2^+(0)$        & $D^*\,D_J^*\vert_P$   & 4511          & 0.577 & $D\,D^*\vert_D$       & 3906          & 0.447 \\
$2^+(1)$        & $D^*\,D^*\vert_S$     & 4040          & 0.470 & $D\,D\vert_D$         & 3772          & 0.424 \\ 
$0^-(0)$        & $D_1\,D_1\vert_P$     & 4910          & 0.664 & $D\,D^*\vert_P$       & 3906          & 0.447 \\ 
$0^-(1)$        & $D_J^*\,D_J^*\vert_P$ & 4982          & 0.684 & $D\,D^*\vert_P$       & 3906          & 0.447 \\ 
$1^-(0)$        & $D_1\,D_J^*\vert_P$   & 4946          & 0.674 & $D\,D\vert_P$         & 3772          & 0.424 \\ 
$1^-(1)$        & $D_1\,D_J^*\vert_P$   & 4946          & 0.674 & $D\,D^*\vert_P$       & 3906          & 0.447 \\ 
$2^-(0,1)$      & $D_J^*\,D_J^*\vert_P$ & 4982          & 0.684 & $D\,D^*\vert_P$       & 3906          & 0.447 \\
\hline
\multicolumn{7}{|c|}{$L=1$ $S=0$}\\
\hline
$1^-(0)$        & $D\,D\vert_P$         & 3772          & 0.424 & $D\,D\vert_P$         & 3772          & 0.424 \\
$1^-(1)$        & $D\,D_1\vert_{S,D}$   & 4341          & 0.544 & $D\,D^*\vert_P$       & 3906          & 0.447 \\
\hline
\multicolumn{7}{|c|}{$L=1$ $S=1$}\\
\hline
$0^-(0,1)$      & $DD^*\vert_P$         & 3906          & 0.447 & $D\,D^*\vert_P$       & 3906          & 0.447 \\
$1^-(0)$        & $D\,D^*\vert_P$       & 3906          & 0.447 & $D\,D\vert_P$         & 3772          & 0.424 \\
$1^-(1)$        & $D\,D^*\vert_P$       & 3906          & 0.447 & $D\,D^*\vert_P$       & 3906          & 0.447 \\
$2^-(0,1)$      & $D\,D^*\vert_P$       & 3906          & 0.447 & $D\,D^*\vert_P$       & 3906          & 0.447 \\ 
\hline
\multicolumn{7}{|c|}{$L=1$ $S=2$}\\
\hline
$1^-(0)$        & $D^*\,D^*\vert_P$     & 4040          & 0.470 & $D\,D\vert_P$         & 3772          & 0.424 \\
$1^-(1)$        & $D^*\,D_J^*\vert_{S,D}$& 4511         & 0.577 & $D\,D^*\vert_P$       & 3906          & 0.447 \\ 
$2^-(0)$        & $D^*\,D^*\vert_P$     & 4040          & 0.470 & $D\,D^*\vert_P$       & 3906          & 0.447 \\
$2^-(1)$        & $D^*\,D_J^*\vert_{S,D}$& 4511         & 0.577 & $D\,D^*\vert_P$       & 3906          & 0.447 \\
$3^-(0)$        & $D^*\,D^*\vert_P$     & 4040          & 0.470 & $D^*\,D^*\vert_P$     & 4040          & 0.470 \\
$3^-(1)$        & $D^*\,D_J^*\vert_{S,D}$& 4511         & 0.577 & $D_1\,D\vert_D$       & 4341          & 0.544 \\ 
\hline
\end{tabular}
\end{table}
\end{center}

\begin{center}
\begin{table}
\caption{Same as Table \ref{th3} for the BCN model.}
\label{th3b}
\begin{tabular}{|c|ccc|ccc|}
\hline
                &\multicolumn{3}{c|}{UN}        &\multicolumn{3}{c|}{CO} \\
\hline
$J^P(I)$        & $M_1\,M_2\vert_L$     &$E(M_1,M_2)$ &RMS      &$M_1\,M_2\vert_L$      & $E(M_1,M_2)$ &RMS\\ 
\hline
\multicolumn{7}{|c|}{$L=0$}\\
\hline
$0^+(0)$        & $B_1\,B\vert_P$       & 11113 & 0.355 & $B_1\,B\vert_P$       & 11113 & 0.355 \\
$0^+(1)$        & $B\,B\vert_S$         & 10602 & 0.282 & $B\,B\vert_S$         & 10602 & 0.282 \\ 
$1^+(0,1)$      & $B\,B^*\vert_S$       & 10651 & 0.288 & $B\,B^*\vert_{S,D}$   & 10651 & 0.288 \\ 
$2^+(0)$        & $B^*\,B_J^*\vert_P$   & 11176 & 0.364 & $B\,B^*\vert_D$       & 10651 & 0.288 \\
$2^+(1)$        & $B^*\,B^*\vert_S$     & 10700 & 0.294 & $B\,B\vert_D$         & 10602 & 0.282 \\ 
$0^-(0)$        & $B_1\,B_1\vert_P$     & 11624 & 0.428 & $B\,B^*\vert_P$       & 10651 & 0.288 \\ 
$0^-(1)$        & $B_J^*\,B_J^*\vert_P$ & 11652 & 0.434 & $B\,B^*\vert_P$       & 10651 & 0.288 \\ 
$1^-(0)$        & $B_1\,B_J^*\vert_P$   & 11638 & 0.431 & $B\,B\vert_P$         & 10602 & 0.282 \\ 
$1^-(1)$        & $B_1\,B_J^*\vert_P$   & 11638 & 0.431 & $B\,B^*\vert_P$       & 10651 & 0.288 \\ 
$2^-(0,1)$      & $B_J^*\,B_J^*\vert_P$ & 11652 & 0.434 & $B\,B^*\vert_P$       & 10651 & 0.288 \\
\hline
\multicolumn{7}{|c|}{$L=1$ $S=0$}\\
\hline
$1^-(0)$        & $B\,B\vert_P$         & 10602 & 0.282 & $B\,B\vert_P$         & 10602 & 0.282 \\
$1^-(1)$        & $B\,B_1\vert_{S,D}$   & 11113 & 0.355 & $B\,B^*\vert_P$       & 10651 & 0.288 \\
\hline
\multicolumn{7}{|c|}{$L=1$ $S=1$}\\
\hline
$0^-(0,1)$      & $BB^*\vert_P$         & 10651 & 0.288 & $B\,B^*\vert_P$       & 10651 & 0.288 \\
$1^-(0)$        & $B\,B^*\vert_P$       & 10651 & 0.288 & $B\,B\vert_P$         & 10602 & 0.282 \\
$1^-(1)$        & $B\,B^*\vert_P$       & 10651 & 0.288 & $B\,B^*\vert_P$       & 10651 & 0.288 \\
$2^-(0,1)$      & $B\,B^*\vert_P$       & 10651 & 0.288 & $B\,B^*\vert_P$       & 10651 & 0.288 \\ 
\hline
\multicolumn{7}{|c|}{$L=1$ $S=2$}\\
\hline
$1^-(0)$        & $B^*\,B^*\vert_P$     & 10700 & 0.294 & $B\,B\vert_P$         & 10602 & 0.282 \\
$1^-(1)$        & $B^*\,B_J^*\vert_{S,D}$&11176 & 0.364 & $B\,B^*\vert_P$       & 10651 & 0.288 \\ 
$2^-(0)$        & $B^*\,B^*\vert_P$     & 10700 & 0.294 & $B\,B^*\vert_P$       & 10651 & 0.288 \\
$2^-(1)$        & $B^*\,B_J^*\vert_{S,D}$&11176 & 0.364 & $B\,B^*\vert_P$       & 10651 & 0.288 \\
$3^-(0)$        & $B^*\,B^*\vert_P$     & 10700 & 0.294 & $B^*\,B^*\vert_P$     & 10700 & 0.294 \\
$3^-(1)$        & $B^*\,B_J^*\vert_{S,D}$&11176 & 0.364 & $B_1\,B\vert_D$       & 11113 & 0.355 \\ 
\hline
\end{tabular}
\end{table}
\end{center}

\begin{center}
\begin{table}[htb]
\caption{Comparison among different numerical approaches to the $cc\bar n \bar n$ system. Energies are in MeV.}
\label{tres-4}
\begin{tabular}{|cccc|}
\hline
$(L,S,I)$ & Ref.~\cite{Vij04b}   & HH($\sum_i \ell_i=0$)& HH\\
\hline
(0,0,1)   & 4155          & 4154  & 3911   \\
(0,1,0)   & 3927          & 3926  & 3860   \\
(0,1,1)   & 4176          & 4175  & 3975   \\
(0,2,1)   & 4195          & 4193  & 4031   \\
\hline
\end{tabular}
\end{table}
\end{center}

\begin{center}
\begin{table}[htb]
\caption{BCN and CQC $(L,S,I)=(0,1,0)$ $cc\bar n \bar n$ results.}
\label{tres-5}
\begin{tabular}{|c|cc|cc|}
\hline
&\multicolumn{2}{c|}{BCN}               &\multicolumn{2}{c|}{CQC}\\
\hline
$K$     & $E_{4q}$ (MeV)        & RMS (fm)      & $E_{4q}$ (MeV)        & RMS (fm)\\ 
\hline
0       & 4100          & 0.310         & 4109          & 0.314 \\ 
2       & 3999          & 0.326         & 3990          & 0.320 \\ 
4       & 3954          & 0.345         & 3931          & 0.331 \\ 
6       & 3933          & 0.364         & 3903          & 0.341 \\ 
8       & 3921          & 0.382         & 3887          & 0.348 \\ 
10      & 3914          & 0.398         & 3878          & 0.354 \\ 
12      & 3910          & 0.414         & 3872          & 0.358 \\ 
14      & 3907          & 0.428         & 3868          & 0.361 \\ 
16      & 3904          & 0.441         & 3866          & 0.363 \\ 
18      & 3903          & 0.453         & 3864          & 0.365 \\ 
20      & 3901          & 0.464         & 3862          & 0.366 \\ 
22      & 3900          & 0.474         & 3861          & 0.367 \\
24      & 3900          & 0.484         & 3861          & $-$   \\
26      & 3899          & 0.492         & $-$           & $-$   \\
\hline
$D\,D^*\vert_S$ & 3906  & 0.447         & 3937          & 0.454  \\
\hline
$\Delta_E$      & \multicolumn{2}{c|}{$-$7}     &\multicolumn{2}{c|}{$-$76} \\
$\Delta_R$      &\multicolumn{2}{c|}{$>1$}      &\multicolumn{2}{c|}{0.808} \\
\hline
\end{tabular}
\end{table}
\end{center}

\begin{center}
\begin{table}[htb]
\caption{$cc\bar n\bar n$ CQC energy, $E_{4q}\equiv E_{4q} (K_{\rm max})$ 
(in MeV), RMS (in fm), $\Delta_R$, and $\Delta_E$ (in MeV) 
as defined in Eqs. (\protect\ref{delta}) and (\protect\ref{delta-r}). 
$M_1\,M_2\vert_L$ indicates the lowest threshold and $E(M_1,M_2)$ its energy as obtained from Table \protect\ref{th2}.}
\label{tres-1}
\begin{tabular}{|cc|cc|cccc|}
\hline
$K_{\rm max}$ & $J^{P}(L,S,I)$ & $M_1\,M_2\vert_L$ & $E(M_1,M_2)$ & $E_{4q}$     & RMS$_{4q}$ & $\Delta_R$ & $\Delta_E$ \\
\hline
28      & $0^+(0,0,0)$  & $D_1\,D\vert_P$       & 4426  & 4441  & 0.624 & $>$1  & +15 \\
28      & $0^+(0,0,1)$  & $D\,D\vert_S$         & 3872  & 3905  & 0.752 & $>$1  & +33 \\
24      & $1^+(0,1,0)$  & $D\,D^*\vert_S$       & 3937  & 3861  & 0.367 & 0.808 & $-$76 \\
24      & $1^+(0,1,1)$  & $D\,D^*\vert_S$       & 3937  & 3972  & 0.779 & $>$1  & +35 \\
30      & $2^+(0,2,0)$  & $D^*\,D_J^*\vert_P$   & 4499  & 4526  & 0.987 & $>$1  & +27 \\
30      & $2^+(0,2,1)$  & $D^*\,D^*\vert_S$     & 4002  & 4025  & 0.879 & $>$1  & +22 \\
&&&&&&&\\
25      & $0^-(0,0,0)$  & $D_1\,D_1\vert_P$     & 4980  & 5374  & 0.738 & $>$1  & +394 \\
25      & $0^-(0,0,1)$  & $D^*_J\,D^*_J\vert_P$ & 4996  & 5012  & 0.982 & $>$1  & +16 \\
25      & $1^-(0,1,0)$  & $D_1\,D_J^*\vert_P$   & 4988  & 5021  & 0.944 & $>$1  & +33 \\
25      & $1^-(0,1,1)$  & $D_1\,D_J^*\vert_P$   & 4988  & 5018  & 0.982 & $>$1  & +30 \\
25      & $2^-(0,2,0)$  & $D_J^*\,D_J^*\vert_P$ & 4996  & 5387  & 1.237 & $>$1  & +391 \\
25      & $2^-(0,2,1)$  & $D_J^*\,D_J^*\vert_P$ & 4996  & 5025  & 0.991 & $>$1  & +29 \\
&&&&&&&\\
21      & $1^-(1,0,0)$  & $D\,D\vert_P$         & 3872  & 3938  & 0.726 & $>$1  & +66 \\
23      & $1^-(1,0,1)$  & $D\,D_1\vert_{S,D}$   & 4426  & 4426  & 0.527 & 0.894\footnotemark\footnotetext{In this
case the radius has still yet to converge for $K=K_{\rm max}$. Its extrapolation gives a value larger than one but finite.} & +0 \\
21      & $(0,1,2)^-(1,1,0)$    & $D\,D^*\vert_P$       & 3937  & 3996  & 0.739 & $>$1  & +59 \\
21      & $(0,1,2)^-(1,1,1)$    & $D\,D^*\vert_P$       & 3937  & 4004  & 0.814 & $>$1  & +67 \\
21      & $(1,2,3)^-(1,2,0)$    & $D^*\,D^*\vert_P$& 4002       & 4052  & 0.817 & $>$1  & +50 \\
19      & $(1,2,3)^-(1,2,1)$    & $D^*\,D_J^*\vert_{S,D}$& 4499& 4461& 0.465&0.766      & $-$38 \\
\hline
\end{tabular}
\end{table}
\end{center}

\begin{center}
\begin{table}[htb]
\caption{$cc\bar n \bar n$ CQC energies and $\Delta_E$, in MeV, evaluated for $K_{\rm max}$ 
and using  the extrapolation (\protect\ref{extra}) in the limit $K\to\infty$.}
\label{tres-2}
\begin{tabular}{|c|cc|cc|}
\hline
$J^P(L,S,I)$    & $E_{4q}(K_{\rm max})$ & $\Delta_E$    & $E_{4q}(K=\infty)$    & $\Delta_E$ \\
\hline
$0^+(0,0,0)$    & 4441  & +15           & 4429  & +3 \\
$0^+(0,0,1)$    & 3905  & +33           & 3862  & $-$10 \\
$1^+(0,1,0)$    & 3861  & $-$76         & 3856  & $-$81 \\
$1^+(0,1,1)$    & 3972  & +35           & 3914  & $-$14 \\
$2^+(0,2,0)$    & 4526  & +27           & 4501  & +2 \\
$2^+(0,2,1)$    & 4024  & +22           & 3991  & $-$9 \\
&&&&\\
$0^-(0,0,0)$    & 5374  & +394          & 5323  & +343 \\
$0^-(0,0,1)$    & 5012  & +16           & 4983  & $-$13 \\
$1^-(0,1,0)$    & 5021  & +33           & 4993  & +5 \\
$1^-(0,1,1)$    & 5018  & +30           & 4992  & +4 \\
$2^-(0,2,0)$    & 5387  & +391          & 5307  & +311 \\
$2^-(0,2,1)$    & 5025  & +29           & 5000  & +4 \\
&&&&\\
$1^-(1,0,0)$    & 3938  & +66           & 3865          &$-$7 \\
$1^-(1,0,1)$    & 4426  & +0            & 4420          &$-$6 \\
$(0,1,2)^-(1,1,0)$ & 3996       & +59           & 3927          &$-$10 \\
$(0,1,2)^-(1,1,1)$ & 4004       & +67           & 3918          &$-$19 \\
$(1,2,3)^-(1,2,0)$ & 4052       & +50           & 3993          &$-$9 \\
$(1,2,3)^-(1,2,1)$ & 4461       & $-$38         & 4461          &$-$38 \\
\hline
\end{tabular}
\end{table}
\end{center}

\begin{center}
\begin{table}[htb]
\caption{Same as Table \ref{tres-1} for the $b b \bar n \bar n$ system.}
\label{tres-3}
\begin{tabular}{|cc|cc|cccc|}
\hline
$K_{\rm max}$ & $J^P(L,S,I)$ & $M_1\,M_2\vert_L$ & $E(M_1,M_2)$ & $E_{4q}$       & RMS$_{4q}$ & $\Delta_R$ & $\Delta_E$ \\
\hline
30      & $0^+(0,0,0)$  & $B\,B_1\vert_P$       & 11101 & 10952 & 0.328 & 0.881 & --149 \\
26      & $0^+(0,0,1)$  & $B\,B\vert_S$         & 10588 & 10606 & 0.365 & $>$1  & +18 \\
22      & $1^+(0,1,0)$  & $B\,B^*\vert_S$       & 10612 & 10398 & 0.220 & 0.765 & $-$214 \\
24      & $1^+(0,1,1)$  & $B\,B^*\vert_S$       & 10612 & 10623 & 0.310 & $>$1  & +11 \\
28      & $2^+(0,2,0)$  & $B^*_J\,B^*\vert_P$   & 11128 & 11144 & 0.627 & $>1$  & +16 \\
30      & $2^+(0,2,1)$  & $B^*\,B^*\vert_S$     & 10635 & 10636 & 0.314 & $>$1  & +1 \\
&&&&&&&\\
23      & $1^-(1,0,0)$  & $B\,B\vert_P$         & 10588 & 10577 & 0.335 & $>$1  & --11 \\
19      & $1^-(1,0,1)$  & $B\,B_1\vert_{S,D}$   & 11101 & 10980 & 0.276 & 0.740 & $-$121\\
21      & $(0,1,2)^-(1,1,0)$    & $B\,B^*\vert_P$       & 10612 & 10650 & 0.493 & $>$1  & +38\\
21      & $(0,1,2)^-(1,1,1)$    & $B\,B^*\vert_P$       & 10612 & 10666 & 0.517 & $>$1  & +54\\
21      & $(1,2,3)^-(1,2,0)$    & $B^*\,B^*\vert_P$& 10635      & 10677 & 0.483 & $>$1  & +42 \\
19      & $(1,2,3)^-(1,2,1)$    & $B^*\,B_J^*\vert_{S,D}$& 11128& 10988& 0.276&0.732    & $-$140 \\
\hline
\end{tabular}
\end{table}
\end{center}

\begin{center}
\begin{table}[htb]
\caption{Same as Table \protect\ref{tres-1} for the BCN model. In all cases the value obtained for
the radius has still yet to converge for $K=K_{\rm max}$.}
\label{tres-3b}
\begin{tabular}{|cc|cc|cccc|}
\hline
$K_{\rm max}$ & $J^P(L,S,I)$ & $M_1\,M_2\vert_L$ & $E(M_1,M_2)$ & $E_{4q}$ & RMS$_{4q}$ & $\Delta_R$ & $\Delta_E$ \\
\hline
26      & $1^+(0,1,0)$  & $D\,D^*\vert_S$       & 3906  & 3899  & 0.492 & $>1$  & $-$7 \\
21      & $1^-(1,0,1)$  & $D\,D_1\vert_{S,D}$   & 4341  & 4380  & 0.640 & $>1$  & +39  \\
21      & $(1,2,3)^-(1,2,1)$& $D^*\,D_J^*\vert_{S,D}$   & 4511  & 4502  & 0.492 & 0.853 & $-$9 \\
\hline
\end{tabular}
\end{table}
\end{center}

\begin{center}
\begin{table}[htb]
\caption{Same as Table \ref{tres-3b} for the $bb \bar n\bar n$ system.}
\label{tres-3c}
\begin{tabular}{|cc|cc|cccc|}
\hline
$K_{\rm max}$ & $J^{P}(L,S,I)$ & $M_1\,M_2\vert_L$ & $E(M_1,M_2)$  & $E_{4q}$    & RMS$_{4q}$ & $\Delta_R$ & $\Delta_E$ \\
\hline
28      & $0^+(0,0,0)$  & $B\,B_1\vert_P$       & 11113 & 11061 & 0.334 & 0.941 & $-$52 \\
28      & $1^+(0,1,0)$  & $B\,B^*\vert_S$       & 10651 & 10507 & 0.220 & 0.764 & $-$144 \\
28      & $2^+(0,2,1)$  & $B^*\,B^*\vert_S$     & 10700 & 10723 & 0.353 & $>$1  & +23 \\
19      & $1^-(1,0,0)$  & $B\,B\vert_P$         & 10602 & 10639 & 0.433 & $>$1  & +27 \\
19      & $1^-(1,0,1)$  & $B\,B_1\vert_{S,D}$   & 11113 & 11037 & 0.265 & 0.746 & $-$76\\
19      & $(1,2,3)^-(1,2,1)$    & $B^*\,B_J^*\vert_{S,D}$& 11176& 11057& 0.264&0.727    & $-$119 \\
\hline
\end{tabular}
\end{table}
\end{center}

\begin{center}
\begin{table}
\caption{Summary of bound states.}
\label{tres-8}
\begin{tabular}{|cccc|}
\hline
Quark content      &  $J^P\,(L,S,I)$      &  Model & Decay mode            \\
\hline
$cc\bar n\bar n$   &  $1^+\,(0,1,0)$      &  CQC   & Weak              \\
                   &                      &  BCN   & Electromagnetic   \\
\hline
$bb\bar n\bar n$   &  $1^+\,(0,1,0)$      &  CQC   &  Weak              \\
                   &                      &  BCN   &  Weak   \\
                   &  $3^-\,(1,2,1)$      &  CQC   &  Electromagnetic   \\
                   &                      &  BCN   &  Electromagnetic   \\
                   &  $0^+\,(0,0,0)$      &  CQC   &  Electromagnetic   \\
                   &                      &  BCN   &  Electromagnetic   \\
                   &  $1^-\,(1,0,0)$      &  CQC   &  Weak              \\
\hline
\end{tabular}
\end{table}
\end{center}

\begin{center}
\begin{table}[htb]
\caption{$\Delta_E$, in MeV, in the uncoupled ($\Delta_E^{\rm UN}$) 
and coupled schemes ($\Delta_E^{\rm CO}$) for different 
states with the CQC and BCN models. See text for details}
\label{tres-7}
\begin{tabular}{|cc|cc|cc|cc|cc|}
\hline
                &       &\multicolumn{4}{|c|}{CQC}&\multicolumn{4}{|c|}{BCN}\\
\hline
                & $J^P (L,S,I)$ & $M_1\,M_2|^{LS}$      & $\Delta_E^{\rm UN}$       & $M_1\,M_2|^J$         & $\Delta_E^{\rm CO}$
                        & $M_1\,M_2|^{J}$       & $\Delta_E^{\rm UN}$       & $M_1\,M_2|^J$         & $\Delta_E^{\rm CO}$ \\
\hline
$cc\bar n\bar n$& $1^- (1,2,1)$ & 4499                  & $-38$         & 3937                  & $+524$ 
                        & 4511                  & $-$9          & 3906                  & $+596$ \\
                & $2^- (1,2,1)$ & 4499                  & $-38$         & 3937                  & $+524$ 
                        & 4511                  & $-$9          & 3906                  & $+596$ \\
                & $3^- (1,2,1)$ & 4499                  & $-38$         & 4426                  & $+35$ 
                        & 4511                  & $-$9          & 4341                  & $+161$ \\
                & $1^- (1,0,1)$ & 4426                  & $0$           & 3937                  & $+489$ 
                        & 4341                  & $+$39         & 3906                  & $+474$ \\
\hline
$bb\bar n\bar n$& $1^- (1,2,1)$ & 11128                 & $-140$        & 10612                 & $+376$
                        & 11176                 & $-119$        & 10651                 & $+406$ \\
                & $2^- (1,2,1)$ & 11128                 & $-140$        & 10612                 & $+376$
                        & 11176                 & $-119$        & 10651                 & $+406$ \\
                & $3^- (1,2,1)$ & 11128                 & $-140$        & 11101                 & $-113$
                        & 11176                 & $-119$        & 11113                 & $-56$ \\
                & $1^- (1,0,1)$ & 11101                 & $-121$        & 10612         & $+368$ 
                        & 11113                 & $-$76                 & 10651         & $+386$ \\
\hline
\end{tabular}
\end{table}
\end{center}

\begin{center}
\begin{table}[htb]
\caption{Probability of the lowest threshold $P_{M_1M_2}$ for different bound and unbound 
four--quark states. Energies in MeV and radii in fm.}
\label{pro}
\begin{tabular}{|ccccccc|}
\hline
Quark content & $J^P(L,S,I)$& $E_{4q}$  & $\Delta_E$    & RMS   & $\Delta_R$    & $P_{M_1M_2}$ \\
\hline
$cc\bar n\bar n$&$0^+(0,0,1)$   & 3877  & +5            & 30.49 & 60.29         & 1.00 \\
                &$1^+(0,1,0)$   & 3861  & $-76$         & 0.37  & 0.81          & 0.50 \\
\hline
$bb\bar n\bar n$&$0^+(0,0,0)$   &10948  & $-$153        & 0.33  & 0.89          & 0.25 \\
                &$1^+(0,1,0)$   &10397  & $-217$        & 0.22  & 0.77          & 0.50 \\
\hline
\end{tabular}
\end{table}
\end{center}

\begin{center}
\begin{table}
\caption{Energies and RMS of the $bb\bar n\bar n$ $2^+(0,2,1)$ 
as a function of $K$ for the CQC model with ($A$) and without ($B$) 
boson exchange potentials.} 
\label{tres-9}
\begin{tabular}{|c|cc|cc|}
\hline
        &\multicolumn{2}{c|}{$A$ }              & \multicolumn{2}{c|}{$B$}\\
\hline
$K$     & $E_{4q}$ (MeV)        & RMS (fm)      & $E_{4q}$ (MeV)        & RMS (fm)      \\ 
\hline
0       & 10763.0       & 0.2131        & 10793.3       & 0.2173 \\
2       & 10701.8       & 0.2201        & 10740.6       & 0.2250 \\
4       & 10670.6       & 0.2283        & 10712.2       & 0.2337 \\
6       & 10658.0       & 0.2360        & 10700.3       & 0.2425 \\
8       & 10650.1       & 0.2436        & 10692.2       & 0.2521 \\
10      & 10645.6       & 0.2509        & 10687.2       & 0.2628 \\
12      & 10642.6       & 0.2580        & 10683.5       & 0.2743 \\
14      & 10640.7       & 0.2647        & 10680.7       & 0.2874 \\
16      & 10639.4       & 0.2712        & 10678.6       & 0.3025 \\
18      & 10638.5       & 0.2775        & 10676.6       & 0.3203 \\
20      & 10637.8       & 0.2836        & 10674.9       & 0.3415 \\ 
22      & 10637.3       & 0.2897        & 10673.3       & 0.3670 \\
24      & 10636.8       & 0.2957        & 10671.7       & 0.3973 \\ 
26      & 10636.5       & 0.3021        & 10670.1       & 0.4325 \\
28      & 10636.2       & 0.3077        & 10668.4       & 0.4720 \\
30      & 10636.0       & 0.3136        & 10666.8       & 0.5146 \\
\hline
$B^*B^*\vert_S$&10635.5&0.2906&10635.5&0.2906\\
\hline
\end{tabular}
\end{table}
\end{center}

\end{document}